\documentclass[aps,twocolumn,a4paper,nofootinbib]{revtex4-1}
\newcommand{\Slash}[1]{{\ooalign{\hfil/\hfil\crcr$#1$}}}
\usepackage[dvipdfmx]{graphicx}
\usepackage{amsmath}

\begin{document}

\title{Symmetry breaking effect on the inhomogeneous chiral transition in the magnetic field}
\author{R. Yoshiike and T. Tatsumi}
\affiliation{Department of physics, Kyoto University, Kyoto 606-8502, Japan}

\begin{abstract}
We study the change of the effect of the current quark mass on the inhomogeneous chiral phase in the QCD phase diagram, and discuss the property of the phase transition by the generalized Ginzburg-Landau expansion. The strong external magnetic field spreads this phase over the low chemical potential region even if the current quark mass is finite. This implies that the existence of this phase can be explored by the lattice QCD simulation.
\end{abstract} 

\maketitle

\section{Introduction}

Exploring the finite density region of the QCD phase diagram is one of the  challenging issues in nuclear physics.
Recently, the possible existence of the inhomogeneous chiral phase has been energetically discussed by the analysis of the Nambu-Jona-Lasinio (NJL) type model \cite{nakano,nickel,ggl} or the Shwinger-Dyson approach \cite{muller}.
In this phase, the scalar and pseudoscalar quark condensates spatially modulate
and the complex order parameter, $\phi({\bf r})$, representing this phase  takes the form,
\begin{align}
 \phi({\bf r}) \equiv  \langle \bar{\psi} \psi \rangle + i \langle \bar{\psi} i\gamma^5 \tau_3 \psi \rangle  = \Delta({\bf r}) e^{i\theta({\bf r})}. \label{condensate} 
\end{align}
As a definite form of $\phi({\bf r})$, the dual chiral density wave (DCDW)  $(\Delta({\bf r}) =\Delta, \theta({\bf r})=qz)$ \cite{nakano,muller}, the real kink crystal (RKC)  $(\Delta({\bf r})\sim \Delta{\rm sn}({\bf \kappa}z),~\theta({\bf r})=0)$ \cite{nickel,ggl} or the hybrid condensate $(\Delta({\bf r})\sim \Delta{\rm sn}({\bf \kappa} z),~\theta({\bf r})=qz)$ \cite{nishiyama} has been often used.
These configurations can be obtained by using the Hartree-Fock solutions in the NJL$_2$ model in the chiral limit \cite{basar,basar2}.
Most of analyses have  shown that the inhomogeneous chiral phase appears as an intermediate phase during the standard chiral phase transition.

Nowadays, various magnetic aspects of QCD have  also attracted much interest 
because there have been expected  quark matter with the strong magnetic field B in early universe, during  the heavy ion collision ($B=\mathcal{O}(m_\pi^2\sim 10^{17}{\rm G})$) or in the core of compact stars ($B>10^{12-15}$G). Any magnitude of $B$ is also possible on the numerical lattice.
One of the interesting subjects is the symmetry behavior in the presence of the magnetic field ($B$).
 It has been suggested that the chiral symmetry breaking is enhanced due to $B$ in the effective model, {\it magnetic catalysis} \cite{suganuma,klevansky,gusynin}.
However, the recent lattice simulations have shown {\it inverse magnetic catalysis} or {\it magnetic inhibition} \cite{bali,endrodi}. This phenomenon is not well understood yet and its origin is still controversial.
It may be plausible that some fluctuation effects become important, since the matgnetic catalysis has been shown within the mean-field approximation.
Recently, to explain this phenomenon within the effective model, the effective four-Fermi coupling constant has been proposed within the NJL model,
where it depends on $B$ through the coupling of the quark or gluon loops with $B$ perturbatively \cite{farias,ferrer} or in the framework of functional renormalization group \cite{mueller,braun}.

In the external magnetic field, DCDW phase is remarkably extended in the  low chemical potential $(\mu)$ region except for $\mu=0$ \cite{frolov}. The energy spectrum of the quark field exhibits the asymmetry,  
which gives rise to such distinctive behavior 
\footnote{In the recent paper we have suggested a possibility of the spontaneous magnetization in DCDW phase due to the spectral asymmetry \cite{yoshiike}.}
\cite{tatsumi}.
Note that complex $\phi({\bf r})$ is necessary for the energy spectrum to be asymmetric about zero.
A peculiar role of the spectral asymmetry can be also seen around the transition point:
it induces a new term in the thermodynamic potential,
and consequently a new Lifshitz point should appear on the $\mu=0$ line in the chiral limit \cite{tatsumi}.
If this is the case, one may expect a direct observation of DCDW by the lattice QCD simulations.
The QCD phase diagram in the finite $\mu$ region has been explored by the lattice QCD simulation,
but its availability is severely restricted due to the sign problem.
Some methods to overcome the sign problem have been proposed:~for example the Taylor expansion method \cite{allton,gavai}, the reweighting method \cite{fodor1,fodor2,fodor3,fodor4},
the canonical approach \cite{alexandru,kratochvila,forcrand3,li},
the analysis of Lee-Yang zero in QCD \cite{barbour,nakamura,nagata}
and the analytic continuation method from imaginary chemical potentials \cite{forcrand1,forcrand2,delia1,delia2,chen}. 
However, these methods are limited in the high temperature ($T$) region, $\mu/T < 1$ region.
Therefore, if the inhomogeneous chiral phase develops in the low $\mu$ region, we may have a chance to observe the existence of this phase by the lattice QCD simulation.

In the present work we shall further discuss this issue in a realistic situation. We study the region around the phase transition by using 
the generalized Ginzburg-Landau (GL) expansion \cite{ggl} with the finite current quark mass.
The current mass is small but should be important below the low energy scale of $\mathcal{O}(10^2)$MeV,
since it is well-known that pion mass of $\mathcal{O}(10^2)$MeV is generated from the tiny current quark mass of several MeV.
Thus it is conceivable that the finite quark mass becomes very important in the vicinity of the critical point,
where the wave number as well as the amplitude becomes very  small.
The current mass explicitly breaks chiral symmetry and the energy degeneracy of states is lost under the symmetry operation;
the degeneracy for $m\rightarrow  -m$ ($Z_2$) in the case of RKC or that for $\theta\rightarrow \theta+\alpha({\rm const})$ ($U(1)$) is lost in the case of DCDW.
Since we must utilize these  states to construct the configuration of  the order parameter together with spontaneously symmetry breaking (SSB),
the current mass is expected to disfavor the appearance of the inhomogeneous chiral phase. 
For RKC the exact solution can be obtained in the massive Gross-Neveu model \cite{schnetz} and
the critical point has been demonstrated to be largely shifted \cite{nickel} to reduce the phase region.
For DCDW, no exact solution is known, but a variational method may work well \cite{karasawa}.
Consequently, the effect of the current quark mass is almost similar to the case of RKC:
the function form of DCDW is largely deformed near the transition point and accordingly the DCDW phase is reduced. 
We shall follow the similar approach here and find the proper solution of $\theta(\bf r)$ instead of $qz$ near the transition point.

In particular, the effect of the finite current quark mass should be important when our idea is confronted with the lattice QCD simulations;
one may also extract more information by changing its value by hand.
We know that the Lifshitz point resides on the $\mu=0$ line in the chiral limit.
Once the current mass is turned on, there arises a competition between the positive effect on the DCDW phase by the magnetic field and the negative effect by the current mass.
Consequently we shall see the critical point should leave the $\mu=0$ line
and some gap is formed between them.
In contrast with the crossover for the usual chiral transition in the presence of the finite current mass,
we shall see that the inhomogeneous transition should still have a clear phase boundary due to the loss of translation symmetry.



The paper is organized as follows:
In Sect.~I\hspace{-.1em}I, we construct the thermodynamic potential by using the generalized GL expansion with the finite current quark mass, 
and the configuration of $\phi({\bf r})$ is determined by the stationary condition.
A peculiar role of the spectral asymmetry of the quark energy eigenvalues are emphasized there.
In Sect.~I\hspace{-.1em}I\hspace{-.1em}I, The phase diagram of the DCDW phase is presented in the presence of the magnetic field
and some features of the phase transition is figured out around the transition point. 
The effect of the inverse magnetic catalysis is discussed there.
The possibility of the observation is also discussed in the lattice QCD simulations, based on Ref.~\cite{kashiwa}, where non-analyticity of the partition function is studied in the DCDW phase.
Sect.~I\hspace{-.1em}V is devoted to summary and concluding remarks.

\section{the thermodynamic potential with finite current quark mass}
The thermodynamic potential near the transition point is given by the generalized GL expansion based on the NJL model \cite{ggl}.
The NJL model Lagrangian takes the form,
\begin{align}
 \mathcal{L}_{\rm NJL} = \bar{\psi}\left( i\Slash{D} - m_c \right)\psi + G\left[ \left( \bar{\psi}\psi \right)^2 + \left( \bar{\psi} i\gamma^5 \tau^a \psi \right)^2 \right], \label{njl}
\end{align}
with the covariant derivative, $D_\mu = \partial_\mu + i\mathcal{Q}A_\mu$,
where $\mathcal{Q}$ is the electric charge matrix in flavor space, $\mathcal{Q}={\rm diag}(e_u, e_d)$, and the $SU(2)$ symmetric quark mass, $m_c \equiv m_u=m_d \simeq 5{\rm MeV}$.
We assume the mean field of the quark condensates,
\begin{align}
 M({\bf r}) \equiv -2G\phi({\bf r}) = me^{i\theta(z)}, \label{cond}
\end{align}
where $m=-2G\Delta$ plays a role of the dynamical quark mass,
and the direction of modulation is taken to be parallel to $z$ axis.
Then, the Lagrangian within the mean field approximation takes the form,
\begin{align}
 \mathcal{L}_{\rm MF} &= \bar{\psi}\left[ i\Slash{D} - m_c - m\left( \cos\theta(z) + i\gamma^5 \tau^3 \sin\theta(z) \right)\right]\psi \notag \\
                               & ~~~~~~~~~~~~~~~~~~~~~~~~~~~~~~~~~~~~~~~~~~~~~~~- \frac{m^2}{4G}. \label{mf}
\end{align}

Taking the external magnetic field $\bf{B}$ along the $z$ axis, the thermodynamic potential can be written up  to the fourth order about the order parameters and its derivative and the first order in $m_c$ as  
\begin{align}
 &\Omega(\mu,T,B) = \Omega_0 \notag \\
 &~~+ \int \frac{d^3{\bf x}}{V} \Big\{ \alpha_1m\cos\theta + \frac{1}{2}\left(\alpha_2+\frac{1}{2G} \right)m^2  + \tilde{\alpha}_2 m\,\left(\sin\theta\right)' \notag \\
                         &~~~~~~~~~~~~~~~+ \frac{\alpha_3}{4} \left[ 4m^3 \cos\theta - m\left( \cos \theta \right)''\right] + \tilde{\alpha}_3 m^2\theta' \notag \\
                         &~~~~~~~~~~~~~~~+ \frac{\alpha_4}{4}\left( m^4 - m^2\theta\theta''\right) + 3\tilde{\alpha}_{4a}m^3\left(\sin\theta\right)' \notag \\
                         &~~~~~~~~~~~~~~~+ \tilde{\alpha}_{4b}m\left(\sin\theta\right)''' \Big\}, \label{omega}
\end{align}
with a shorthand notation, $\theta'\equiv \partial\theta/\partial z$, for given $\mu,~T$ and $B$. The GL coefficients read,
\begin{align}
 &\alpha_{2j} = (-1)^j 2N_c\sum_f T\sum_{k}\frac{|e_fB|}{2\pi} \sum_{n\geq0} \notag \\
 &~~~~\times \int \frac{dp}{2\pi}\frac{2-\delta_{n,0}}{\left[ (\omega_k + i\mu)^2 + p^2 + 2|e_fB|n \right]^j}, \label{a2m} \\
 &\alpha_{2j-1} = m_c \alpha_{2j}, \label{a2m-1} \\
 &\tilde{\alpha}_3 = N_c\sum_f\frac{|e_fB|}{16\pi^3 T} {\rm Im}\psi^{(1)}\left(\frac{1}{2}+i\frac{\mu}{2\pi T}\right), \label{a3} \\
 &\tilde{\alpha}_2 = m_c\tilde{\alpha}_3, \\
 &\tilde{\alpha}_{4b} = 
m_cN_c\sum_f\frac{|e_fB|}{1536\pi^5 T^3} {\rm Im}\psi^{(3)}\left(\frac{1}{2}+i\frac{\mu}{2\pi T}\right),
\end{align}
where $\omega_k=(2k+1)\pi T$ is the Matsubara frequency and $\Omega_0$ is the constant term independent of the order parameters.
The derivation of these equations is somewhat cumbersome and is relegated to Appendix A.
Here $\tilde{\alpha}_{4a}$ cannot be represented as a simple form (see Appendix A for details).
Note that the effect of the current quark mass appears in $\alpha_{2j-1}$, $\tilde{\alpha}_2$, $\tilde{\alpha}_{4a}$ and $\tilde{\alpha}_{4b}$, which are proportional to $m_c$.
The coefficients $\alpha_i~(i=1-4)$ include a ultraviolet divergence and should be properly regularized by applying some regularization scheme.
In the present calculation, the Pauli-Villars regularization (PVR) is used (Appendix B).

It may be worth mentioning that the $\tilde{\alpha}_3$ term is originated from the spectral asymmetry of the quark energy eigenvalues and proportional to $B$. The presence of such term has been shown in the chiral limit 
and a close relation to chiral anomaly has been demonstrated \cite{tatsumi}. This argument can be easily generalized even if the current mass is taken into account (see Appendix C).
Note that  $\tilde{\alpha}_3$ term remarkably extends the DCDW phase in the presence of the magnetic field \cite{tatsumi}, while it cannot appear in the RKC phase because of the absence of the phase degree of freedom. 

\begin{figure}[t]
 \centering
     \begin{center}
        \includegraphics[width=8.5cm]{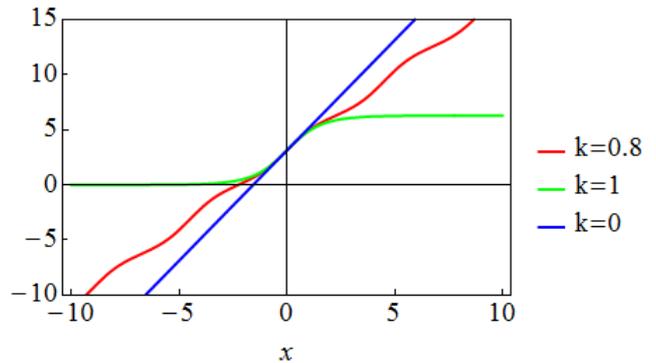}
      \end{center}
 \vspace{-0.5cm}
 \caption{Plot of $\pi + 2{\rm am}(x,k)$. The red, green and blue line describes the function at $k=0.8,\,1,\,0$ respectively.}
 \label{theta}
\end{figure}

The surface terms in Eq.~(\ref{omega}) are irrelevant for the stationary condition: $\delta \Omega/\delta\theta(z)=0$.
Thus, we find the equation in the sine-Gordon form,
\begin{align}
 \theta'' + {\rm sign}(\alpha_1 + m^2\alpha_3)m^{*2}_\pi\sin\theta = 0, \label{sine}
\end{align}
with,
\begin{align}
 m_\pi^{*2} \equiv 2\frac{|\alpha_1 + m^2\alpha_3|}{m\alpha_4},
\end{align}
and the relevant solution to Eq.~(\ref{sine}) can be obtained as,
\begin{align}
 \theta(z) =2{\rm am}\left( \frac{m^*_\pi}{k} z, k \right) + \pi\theta \left(-\alpha_1 - m^2\alpha_3\right), \label{sol}
\end{align}
where ``am'' is the amplitude function with modulus $k \in [0,1]$.
Note here that the $\tilde{\alpha}_3$ term never affects the stationary condition; it plays instead an important role through the thermodynamic potential.
Then, the period $(l)$ and the wave number $(Q)$ of condensates are defined by the relations,
\begin{align}
 l = \frac{2kK(k)}{m^*_\pi},~Q = \frac{2\pi}{l} = \frac{\pi m^*_\pi}{kK(k)},
\end{align}
where $K(k)$ and $E(k)$ are the complete elliptic integrals.
There are two order parameters, $m$ and $k$ (or $Q$), where 
$m$ characterizes the magnitude of SSB, and 
$k$ measures a degree of the inhomogeneity.
We plot the function: $\pi+ 2{\rm am}(x,k)$ in Fig.~\ref{theta}.
When $k=1$, Eq.~(\ref{sol}) takes the form,
\begin{align}
 \theta(z)|_{k=1} = 4\tan^{-1}\left( e^{m^*_\pi z} \right) - \pi\theta \left(\alpha_1 + m^2\alpha_3\right),
\end{align}
and behaves like the single kink.
Accordingly, $l$ diverges and $Q$ vanishes because $K(k\rightarrow1)\rightarrow \infty$.
Then, we can see that the thermodynamic potential is reduced to the one in the homogeneous phase.
On the other hand, when $k$ and $m_c$ simultaneously go to zero and $2m^*_\pi/k \rightarrow q$,
Eq. (\ref{sol}) takes the form,
\begin{align}
 \theta(z) \rightarrow qz + \pi\theta \left(-\alpha_1 - m^2\alpha_3\right),
\end{align}
and the original DCDW phase is recovered.
In the following, we call the phase where $0<k<1,m\neq0$ the {\it massive DCDW phase}. 

Then the thermodynamic potential takes the form,
\begin{align}
 \Omega &= \Omega_0 - \left|\alpha_1m + \alpha_3 m^3 \right| C_1(k) + \frac{1}{2}\left( \alpha_2 + \frac{1}{2G}\right) m^2 \notag \\
             &~~~~+ \tilde{\alpha}_3 \sqrt{2\frac{|\alpha_1 + m^2\alpha_3|}{\alpha_4}}m^{3/2}C_3(k) + \frac{\alpha_4}{4}m^4, \label{pot}
\end{align}
with
\begin{align}
 C_1(k) 
         &\equiv \frac{2}{k^2} - 1 - \frac{4E (k)}{k^2K(k)}, \\
 C_3(k) 
         &\equiv \frac{\pi}{kK(k)}.
\end{align}
Note that $\tilde{\alpha}_2,\tilde{\alpha}_{4a},\tilde{\alpha}_{4b}$ terms vanish by the spatial integral.
We can easily observe that Eq.~(\ref{pot}) restored the thermodynamic potential in the homogeneous phase at $k\rightarrow1$ because $C_1(k\rightarrow1)=1$ and $C_3(k\rightarrow1)=0$.

One may also find another possible solution of Eq.~(\ref{sine}),
\begin{align}
 \theta_{\rm os}(z) = 2\cos^{-1}\left[k'\,{\rm sn}\left( m^*_\pi z,k' \right) \right] + \pi\theta \left(-\alpha_1 - m^2\alpha_3\right), \label{os}
\end{align}
where ``sn'' is the Jacobi elliptic function with modulus $k' \in [0,1]$.
The previous solution (\ref{sol}) is the monotonically increasing function while this solution is the oscillating function.
Then the thermodynamic potential takes the form,
\begin{align}
 \Omega =& \Omega_0 - \left| \alpha_1m + \alpha_3 m^3 \right| C^{\rm os}_1(k') + \frac{1}{2}\left( \alpha_2 + \frac{1}{2G}\right) m^2 \notag \\
              &+ \frac{\alpha_4}{4} m^4,
\end{align}
with
\begin{align}
 C^{\rm os}_1(k') \equiv 3 - 2k^2 - \frac{4E (k')}{K(k')}.
\end{align}
When $k'=1$, the solution (\ref{os}) corresponds to $\theta(z)|_{k=1}$
and the thermodynamic potential becomes the one in the homogeneous phase.
However, we can see that the oscillating solution is never favored compared to the homogeneous solution
because $C^{\rm os}_1(k') \leq C^{\rm os}_1(k'=1)$.
Therefore, the phase with the oscillating solution does not appear in the present situation
\footnote{The oscillating solution may be relevant near the critical point in the absence of the magnetic field, where 
the similar equation is derived for $\theta$ \cite{karasawa}.
}.

\begin{figure*}[ht]
 \centering
   \begin{tabular}{c}
     \begin{minipage}{0.6\hsize}
        \begin{center}
          \includegraphics[width=8.5cm]{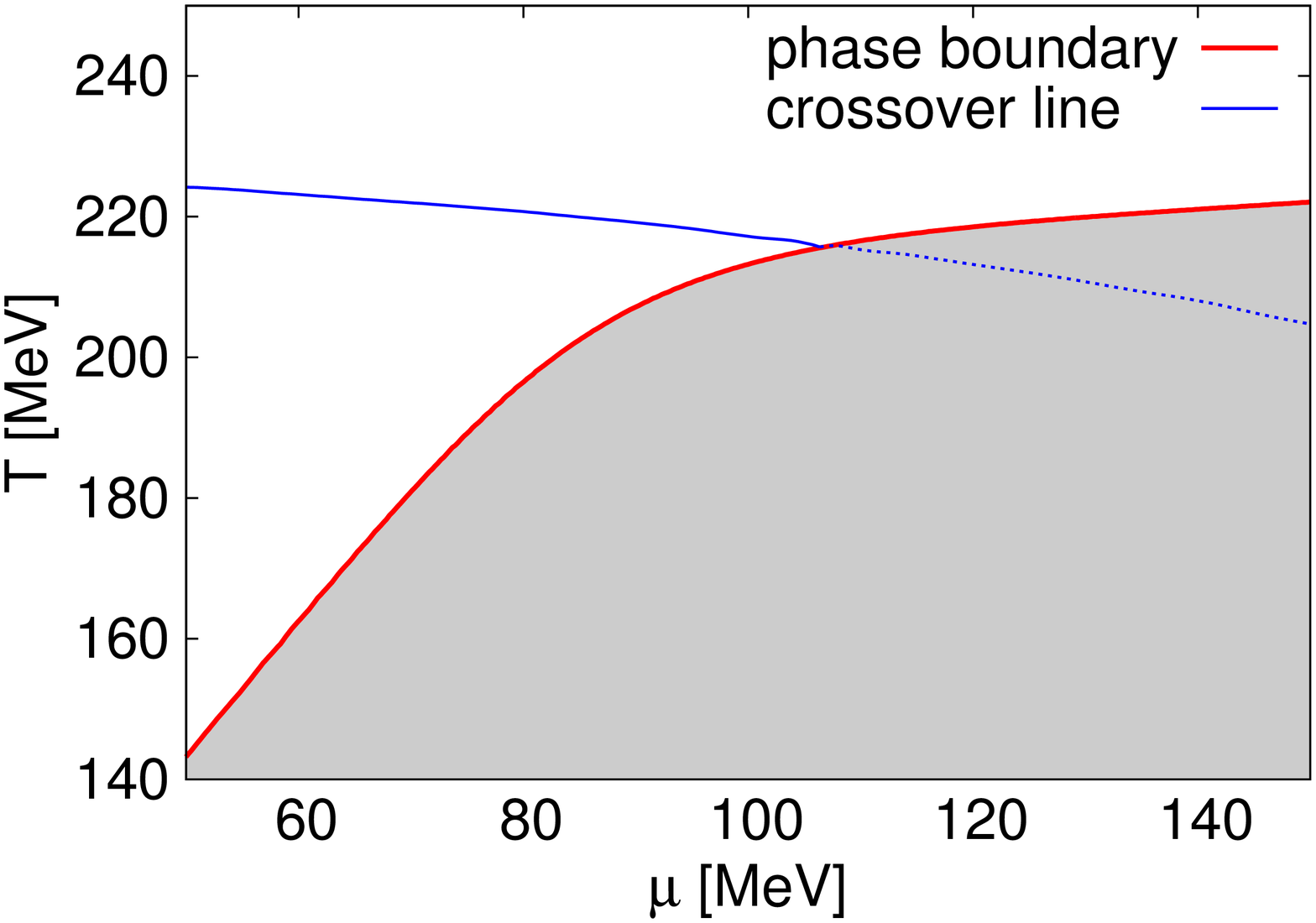}
        \end{center}
     \end{minipage}

     \begin{minipage}{0.4\hsize}
        \begin{center}
          \includegraphics[width=5cm]{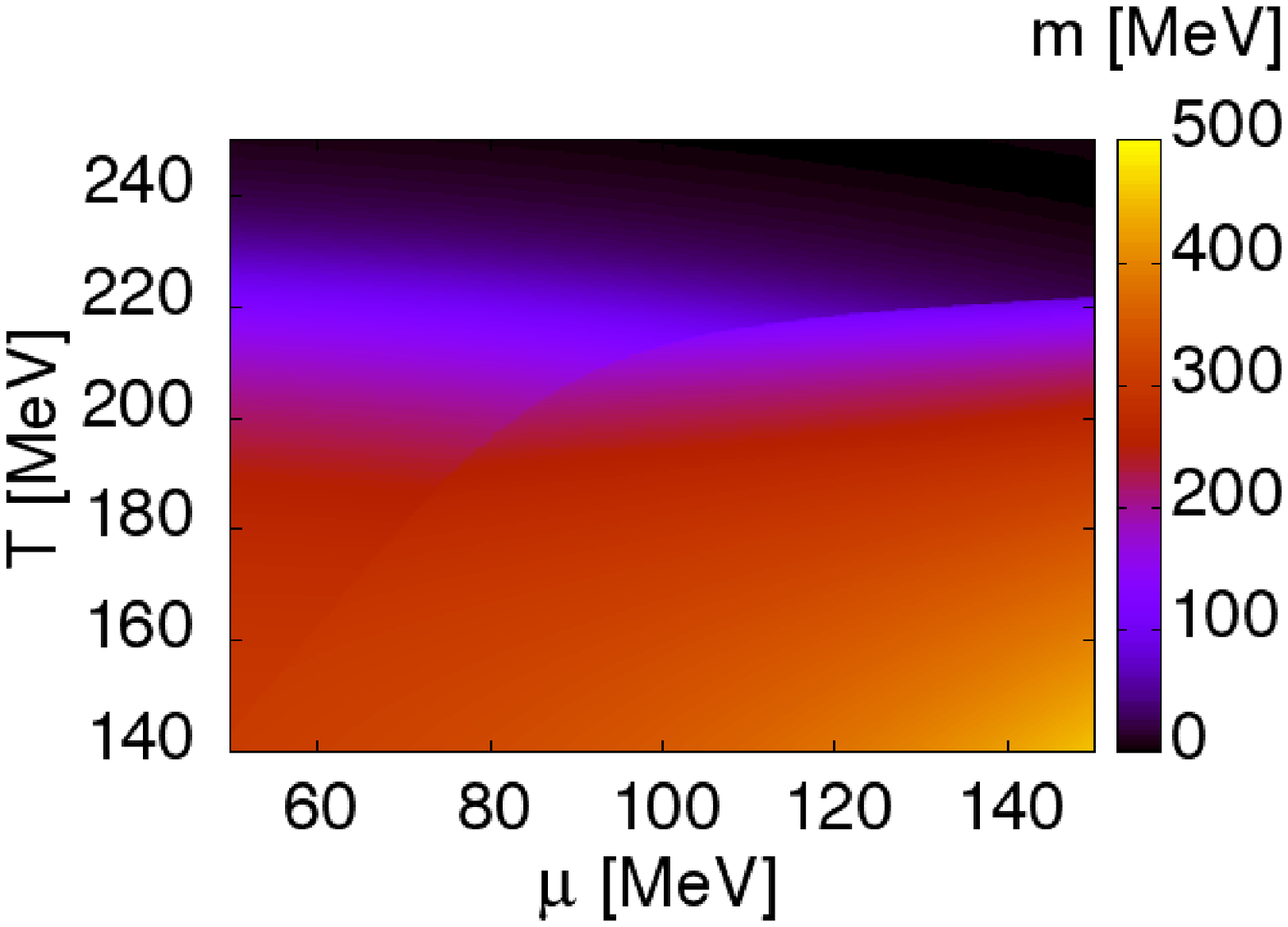}
        \end{center}
        \vspace{-0.7cm}
        \begin{center}
          \includegraphics[width=5cm]{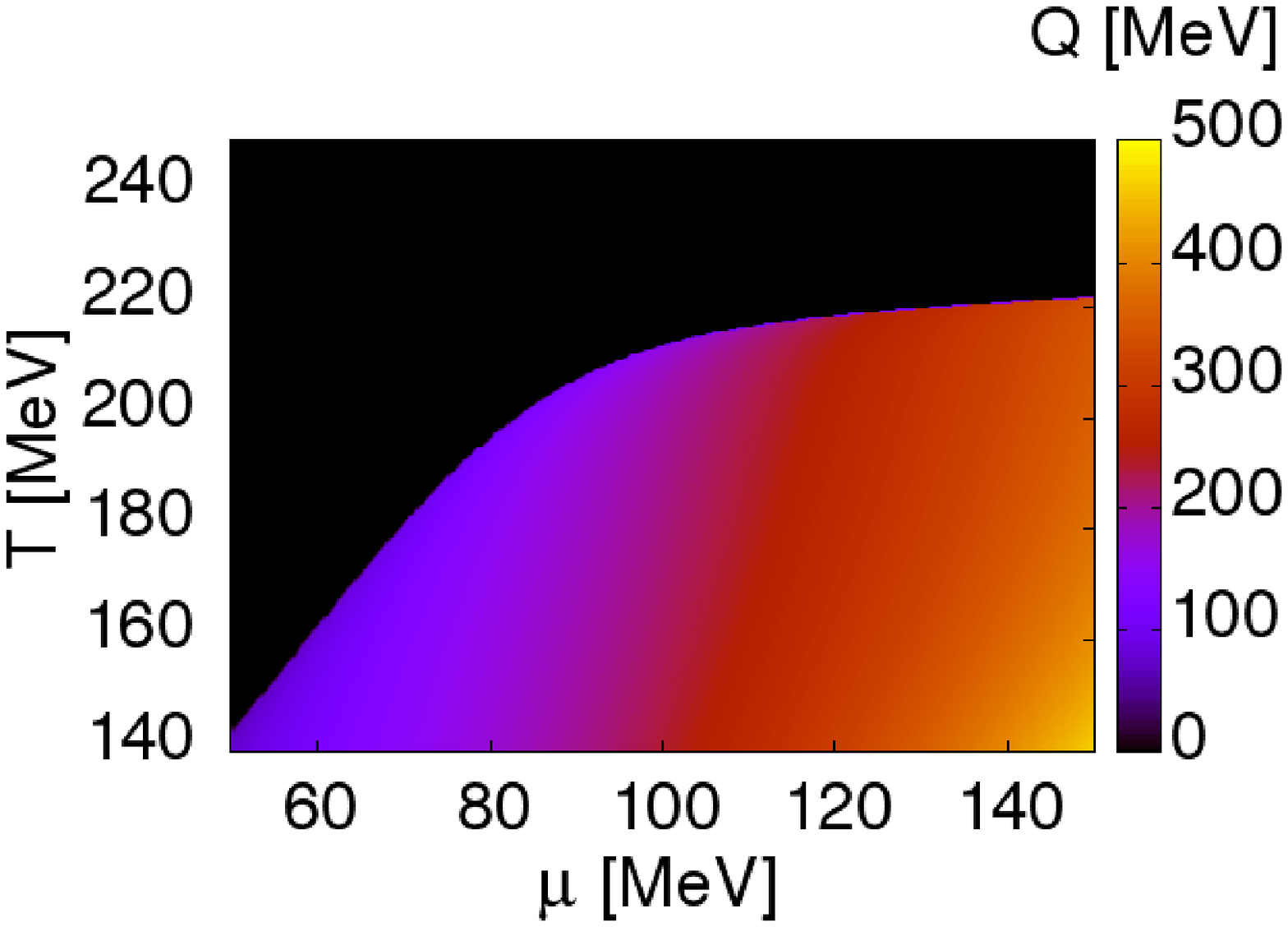}
        \end{center}
      \end{minipage}      
 \end{tabular}
 \caption{Phase diagram at $m_c=5{\rm MeV},\sqrt{eB}=1{\rm GeV}$ (left panel). The red line describes the phase boundary between the massive DCDW phase (shaded area) and the homogeneous phase. The solid blue line describes the crossover line. The conventional crossover line without the massive DCDW phase corresponds to the dashed blue line. The right upper (lower) panel shows the value of $m$ $(Q)$ at the same range of $\mu-T$ as the left panel.}
 \label{diagram}
\end{figure*}

\begin{figure*}[ht]
 \centering
 \begin{tabular}{c}
      \begin{minipage}{0.5\hsize}
        \begin{center}
          \includegraphics[width=8.5cm]{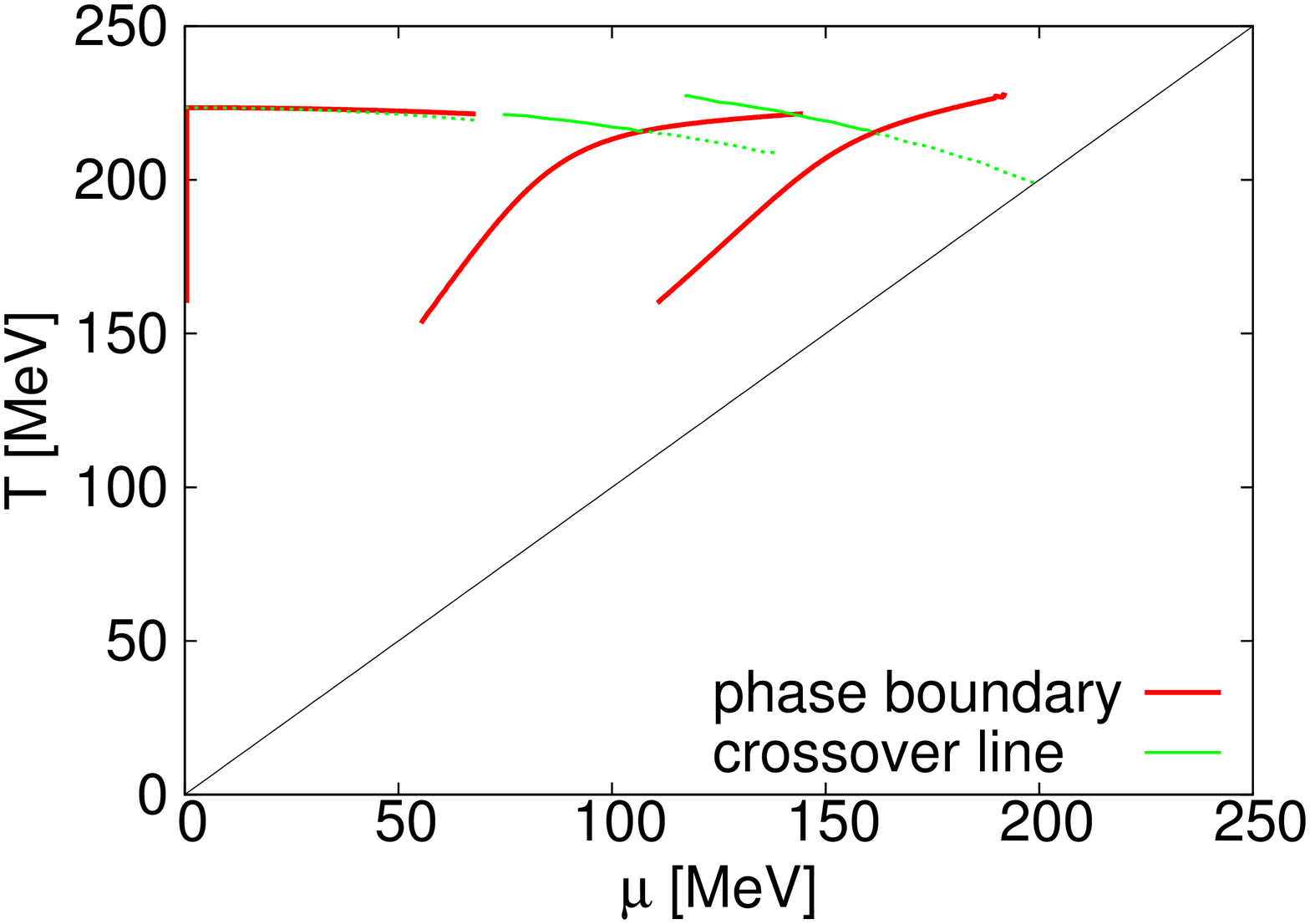}
        \end{center}
      \end{minipage}

      \begin{minipage}{0.5\hsize}
        \begin{center}
          \includegraphics[width=8.5cm]{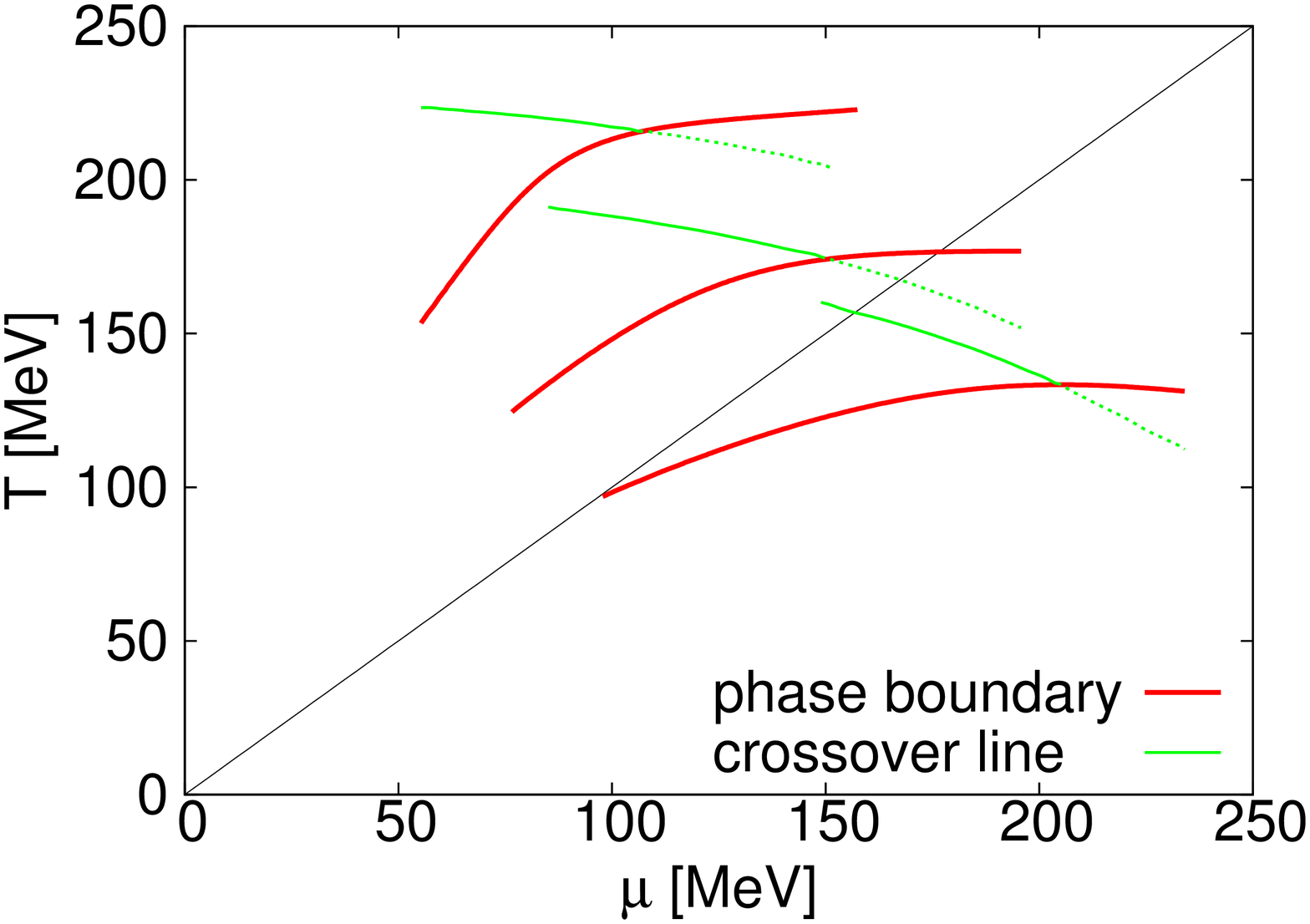}
        \end{center}
      \end{minipage}      
 \end{tabular}
 \vspace{0.5cm}
 \caption{Change of the phase boundary. The left panel shows the result at $m_c=0,\,5,\,20{\rm MeV}$ and fixed $\sqrt{eB}=1{\rm GeV}$. The right panel shows the result at $\sqrt{eB} = 0.5,\,0.7,\,1{\rm GeV}$ and fixed $m_c=5{\rm MeV}$. The red line describes the phase boundary between the massive DCDW phase and the homogeneous phase. The green line describes the crossover line.}
 \label{change}
\end{figure*}

\section{Results and discussions}




\begin{figure*}[t]
 \centering
 \begin{tabular}{c}
      \begin{minipage}{0.5\hsize}
        \begin{center}
          \includegraphics[width=8.5cm]{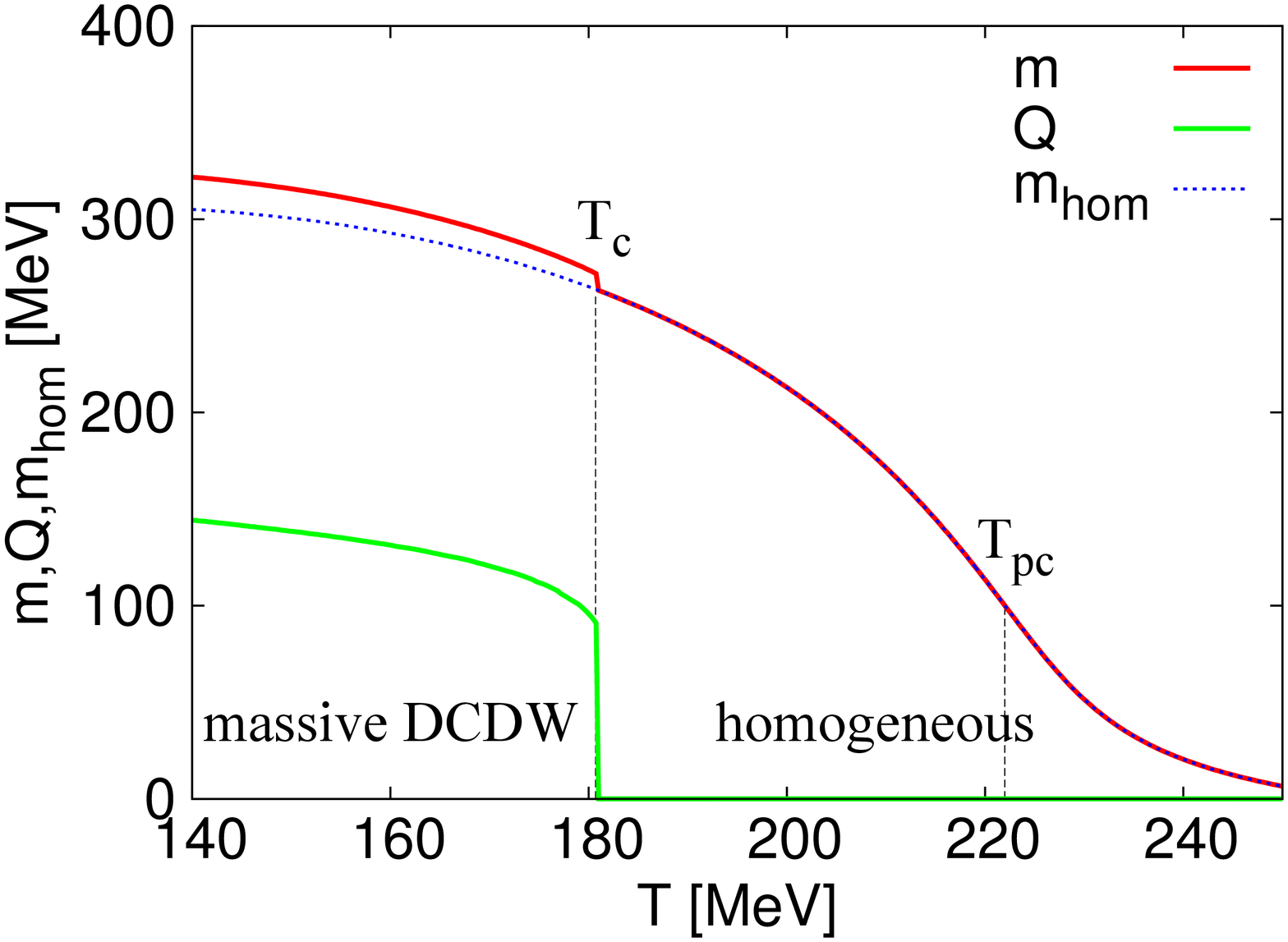}
        \end{center}
      \end{minipage}

      \begin{minipage}{0.5\hsize}
        \begin{center}
          \includegraphics[width=8.5cm]{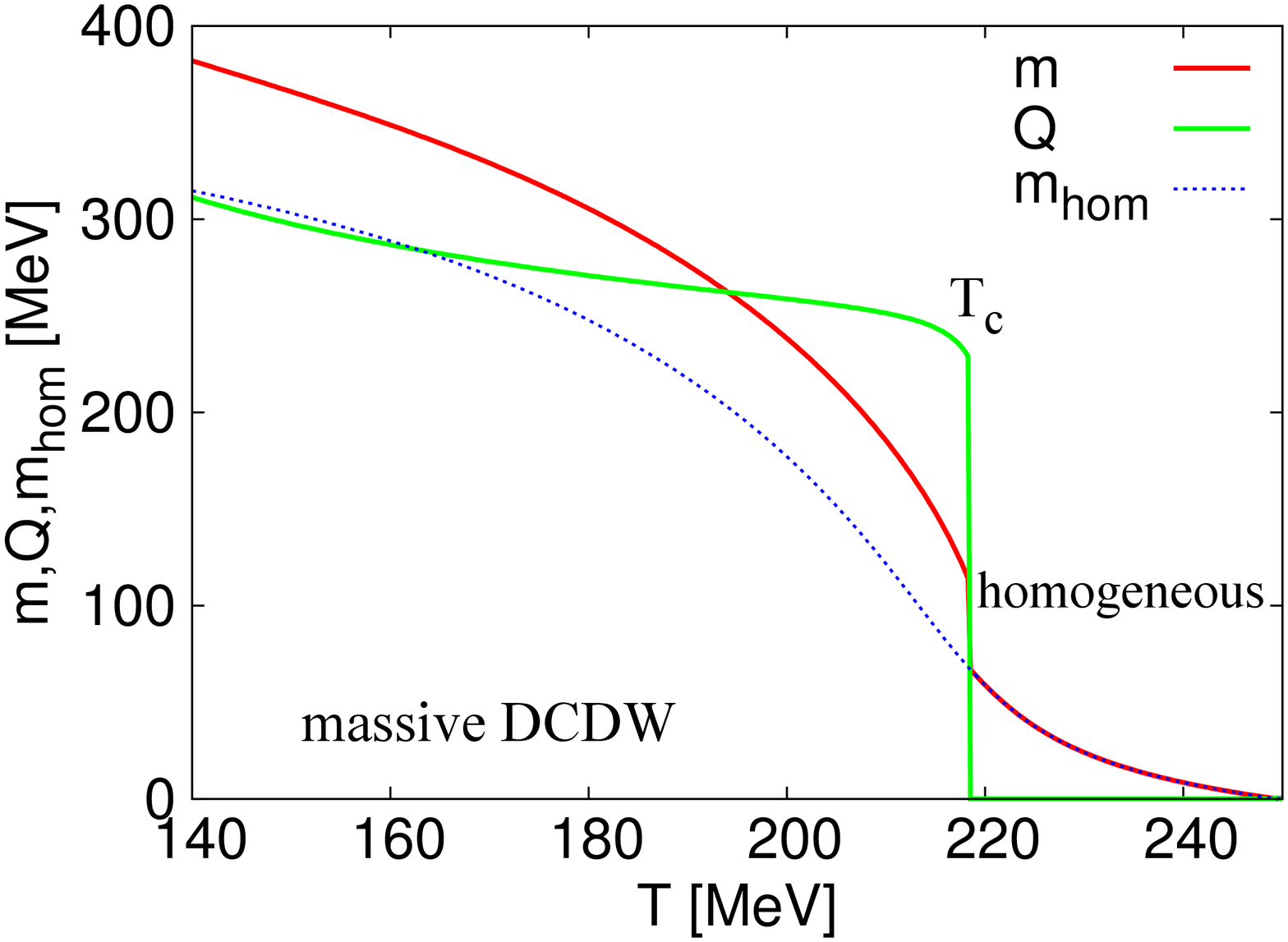}
        \end{center}
      \end{minipage}      
 \end{tabular}
 \vspace{0.5cm}
 \caption{Dependence of the order parameters on $T$ for the same parameter set in Fig. \ref{diagram}. The red or green line describes the amplitude $m$ or the wave number $Q$ respectively. The dashed blue line shows the conventional dynamical quark mass without the inhomogeneous chiral condensate. The left panel shows the result at $\mu=70{\rm MeV}$ and there are the phase transition point between the homogeneous phase and the massive DCDW phase on $T_c=181{\rm MeV}$ and the pseudocritical point on $T_{pc}=222{\rm MeV}$. The right panel shows the result at $\mu = 120{\rm MeV}$ and there is the phase transition point on $T_c=219{\rm MeV}$.}
 \label{parameter}
\end{figure*}



\subsection{Phase diagram around the transition point}

For obtaining the phase diagram, the order parameters are determined to minimize Eq.~(\ref{pot}).
In the following, $Q$ is used as the order parameter characterizing the inhomogenity instead of $k$.
In the present calculation,  we use the parameter set in Ref.~\cite{njl}:~$\Lambda = 851 {\rm MeV}$ and $G\Lambda^2 = 2.87$, which reproduce pion decay constant $f_\pi=93{\rm MeV}$, pion mass $m_\pi=135{\rm MeV}$ and scalar condensate $\langle \bar{\psi}\psi \rangle=(-250{\rm MeV})^3$ in the vacuum with $m_c=5.2{\rm MeV}$.

In Fig.~\ref{diagram}, we show the resulting phase diagram at $m_c=5{\rm MeV},\sqrt{eB} = 1{\rm GeV}$.
There are the phase boundary  between the massive DCDW phase and the homogeneous phase and the crossover line constituted by the pseudocritical temperature $(T_{pc})$ defined as the peak of the chiral susceptibility:~$-\partial m/\partial T$.

In Fig.~\ref{change}, the change of the phase diagram is described when $m_c$ or $B$ changes.
We can find out that the massive DCDW phase is extended to the low $\mu$ region with the decrease of $m_c$.
Then the result in Ref.~\cite{tatsumi} is recovered in the chiral limit:~$m_c=0$
and it is expected that the crossing point of the phase boundary and the crossover line agrees with the LP in the chiral limit.
On the other hand, $B$ raises the critical temperature in the phase transition, which is consistent with the magnetic catalysis.
In other words, the smaller $m_c$ or the larger $B$ becomes,
the more widely the massive DCDW phase develops over the region:~$\mu/T<1$.

The dependence of the order parameters on $T$ is shown in Fig.~\ref{parameter}.
The discontinuity in the both order parameters can be found at the critical temperature $(T_c)$.
Therefore, it can be concluded that there is the first order phase transition between the massive DCDW phase ($m$ is large and $Q\neq0$) and the homogeneous phase ($m$ is small but finite and $Q=0$)
though there is the second order phase transition between the DCDW phase and the chiral restored phase in the chiral limit \cite{tatsumi}.
The difference can be understood by the fact that $Q$ never becomes redundant because the chiral symmetry is always broken due to the finite $m_c$.
In the right panel, we can see that the first order phase transition is strong
while it becomes weaker at the lower $\mu$.
The crossover between the homogeneously chiral-broken phase and the nearly-restored phase is also observed at $T=T_{\rm pc}$ in the left panel.
The RKC or DCDW phase appears in the region: $\mu\gtrsim300{\rm MeV}$ and $T\lesssim50{\rm MeV}$ with $m_c$ and $B=0$ \cite{nickel,karasawa}.
However we can see that $B$ enlarges the massive DCDW phase over the low $\mu$ and high $T$ region even if $m_c$ is finite.
Furthermore the dynamical quark mass in the massive DCDW phase is larger than the conventional one 
and they correspond after the phase transition.
In other words, the chiral symmetry breaking is promoted in the massive DCDW phase. 
It maybe consistent with the result in the chiral limit \cite{tatsumi}.

\subsection{Effect of the inverse magnetic catalysis}

\begin{figure}[t]
 \centering
        \begin{center}
          \includegraphics[width=8.5cm]{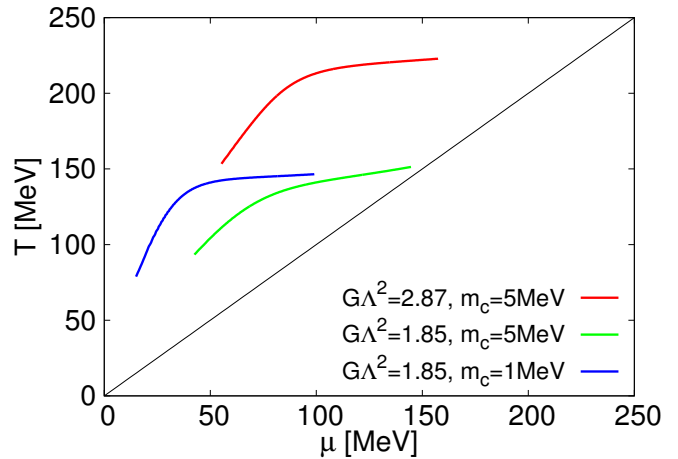}
        \end{center}
 \caption{Phase boundary considered including the inverse magnetic catalysis. The red line corresponds to the phase boundary in the Fig. \ref{diagram}. On the other hand, the green and blue lines describe one at $m_c=5,\,1{\rm MeV}$ with the inverse magnetic catalysis.}
 \label{catal}
\end{figure}

In this subsection, the effect of the inverse magnetic catalysis is discussed in the present model.
Here, it is assumed that the effect is described by giving the $B$ dependence to the coupling constant of the NJL model $(G)$. 
According to Ref.~\cite{ferreira}, $G$ is fitted as reproducing the result of the lattice simulation \cite{bali,endrodi}.
At the parameter set:~$\Lambda = 851 {\rm MeV},~G\Lambda^2 = 2.87,~m_c=5{\rm MeV}$, $T_{pc}(eB=0) = 173{\rm MeV}$ at $\mu=0$.
In the following, we consider the case at $\sqrt{eB}=1{\rm GeV}$.
The coupling constant is putted as $G\Lambda^2=1.85$, which gives the ratio: $T_{pc}/T_{pc}(eB=0)=0.86$ at $\mu=0$.
In Fig.~\ref{catal}, the change of the phase boundary by the inverse magnetic catalysis is shown.
The region of the massive DCDW phase shrinks and the phase transition temperature decreases due to the effect.
However, the massive DCDW phase remains in the $\mu/T<1$ region if $m_c$ is sufficiently small.

\subsection{Possibility of the observation of the inhomogeneous chiral phase}
In Ref.~\cite{kashiwa}, the case with the singular line at $\mu=0$ is discussed.
Though the existence of the line is pointed out by the generalized GL expansion with $B$ in the chiral limit \cite{tatsumi},
such a singular line, that is, the phase boundary is moved to $\mu\neq0$ region due to the current quark mass.
The discussion becomes somewhat simple in this case.
In the Taylor expansion method, some quantity is expanded around $\mu/T=0$ for considering the effect of the finite $\mu$.
Therefore, this method cannot describe the singularity at $\mu\neq0$
and the massive DCDW phase cannot be grasped.
For the same reason, the analytic continuation method from imaginary chemical potential to real chemical potential does not work either.
In other words, the applicable region of these methods is extremely restricted for the massive DCDW phase. 

The reweighting method can overcome the difficulty of the singularity in principle.
In this method, the importance sampling is carried out for some parameter choice, for example Re$\,\mu=0$, where there is no sign problem.
However, the massive DCDW phase does not develop in that region.
Therefore we need to find a special region with the massive DCDW phase and no sign problem there.

In the canonical approach, we also do not have a trouble of the singularity
though the grand canonical potential with the real $\mu$ can be constructed from the one with the imaginary $\mu$.
If there is the massive DCDW phase in $\mu\neq0$ region,
it may be found that the quark number density has the discontinuity derived from some first order phase transition.
However, the phase transition cannot be identified as one from the homogeneous phase to the massive DCDW phase.
Therefore we need to find some specific order parameters on the phase transition.
There is a similar difficulty in the Lee-Yang zero analysis in QCD.
The behavior of zeros of the partition function indicates the existence of some phase transition.
However, we cannot distinguish the phase transition including the massive DCDW phase by their distribution.

We also comment on the two color lattice QCD (QC$_2$D).
In the QC$_2$D, there is no sign problem because the quark determinant is always real even if the chemical potential is real and finite \cite{kogut}.
Therefore, the existence of the inhomogeneous chiral phase may be investigated by the usual Monte Carlo simulation.
It is also thought that this analysis works without sufficiently small $m_c$.
It will be discussed elsewhere \cite{kashiwa2}.

\section{summary and concluding remarks}
We have discussed the inhomogeneous chiral phase at $B\neq0$ and $m_c\neq0$.
In this paper, the thermodynamic potential around the phase transition is obtained by the generalized GL expansion based on the NJL model.
It is found that $B$ extends the massive DCDW phase over the low $\mu$ region similar to the DCDW phase in the chiral limit
though $m_c$ tends to reduce this phase.
Then, there is the first order phase transition between the massive DCDW phase and the homogeneous phase.
Furthermore, the chiral symmetry is strongly broken in this phase compared to the conventional homogeneous phase.

Within our analysis based on the NJL model, $B$ seems to raise the critical temperature.
A similar mechanism to the magnetic catalysis should lead to this behaviour.   
So we adjust the coupling constant of the NJL model to estimate the qualitative influence of the inverse magnetic catalysis.
As a consequence, the critical temperature decreases.
However, the massive DCDW phase can develop in the region:~$\mu/T<1$
if $m_c$ is sufficiently small.
Therefore we suggest that the inhomogeneous chiral phase can be found by the lattice QCD simulations just by choosing some proper method, for example the reweighting method or the canonical approach.
Since there is little work where the local chiral condensate is discussed \cite{iritani},
it is a challenging work to actually confirm the existence of the inhomogeneous chiral phase by the lattice QCD simulations.

On the other hand, the possibility of the massive DCDW phase in $B$ is also interesting from the view of the phenomenology.
It is thought that the quark matter including s-quarks exists with strong magnetic field in neutron stars.
Therefore the phase structure of massive quark matter is needed to discuss properties of neutron stars.
Though they assume the s-quark condensate homogeneous in the previous works \cite{moreira,eos},
s-quarks may be inhomogeneously condensed in neutron stars.
Since the analysis in this paper works only at high temperature,
we need investigate the growth of the massive DCDW phase at zero or low temperature.

\begin{acknowledgments}
The authors thank K. Nishiyama for collaboration in the initial stage of this work. We also thank K. Kashiwa for useful discussions and comments.

This work is partially supported by Grants-in-Aid for Japan Society for the Promotion of Science (JSPS) fellows No. 27-1814 and Grants-in-Aid for Scientific Research on Innovative Areas through No. 24105008 provided by MEXT.
\end{acknowledgments}

\appendix
\section{Generalized GL expansion of the thermodynamic potential}
In this appendix, the thermodynamic potential is expanded about the order parameter and its derivative with $m_c\neq0$ and the external magnetic field $(B)$ along the $z$ axis based on Nickel's work \cite{ggl}.
The thermodynamic potential of the NJL model in the mean field approximation takes the form,
\begin{align}
 &\Omega{(\mu,T,B)} \notag \\
 &= -\frac{T}{V} {\rm{Tr}}_{D,c,f,V} \, {\rm{Ln}}\left[S_B^{-1} - \left( {\rm Re}\tilde{M} + i\gamma^5 \tau^3 {\rm Im}\tilde{M} \right) \right] + \frac{|M|^2}{4G} \notag \\
 &= \Omega_0 -\frac{T}{V} \sum_{j\geq1}\frac{1}{j} {\rm{Tr}}_{D,c,f,V} \left[S_B\left( {\rm Re}\tilde{M} + i\gamma^5 \tau^3 {\rm Im}\tilde{M} \right) \right]^j \notag \\
  &~~~~~~~~~~~~~~~~~~~~~~~~~~~~~~~~~~~~~~~~~~~~~~~~~~~+ \frac{|M|^2}{4G}, \label{ggl}
\end{align}
with $\tilde{M}\equiv m_c+M(z)$, where $\Omega_0$ is independent on the order parameters.
$S_B$ corresponds to the propagator in the chiral limit,
\begin{align}
 S_B = \frac{1}{i\Slash{D}+\mu\gamma^0}. \label{prop}
\end{align}
Then odd $j$ parts always vanish by the Dirac trace.
We need the expansion up to the fifth order about $\tilde{M}$ and its derivative to obtain the thermodynamic potential constituted by the terms up to the fourth order about $M$ and its derivative and the first order in $m_c$.
The thermodynamic potential is expanded into the form in $B=0$ \cite{kamikado},
\begin{align}
 \Omega =& \Omega_0 + \int \frac{d^3{\bf x}}{V} \notag \\
               & \times \left\{ \frac{\alpha_2}{2} |\tilde{M}|^2 + \frac{\alpha_4}{4} \left[|\tilde{M}|^4 - {\rm Re}\left(\tilde{M}\tilde{M}''\right) \right] +\frac{|M|^2}{4G}\right\} \notag \\
             =& \Omega_0 + \int \frac{d^3{\bf x}}{V} \bigg[ \frac{\alpha_2}{2} \left( |M|^2 + 2m_c{\rm Re}M \right) \notag \\
               &~~~+ \frac{\alpha_4}{4} \left(|M|^4 + 4m_c|M|^2{\rm Re}M + |M'|^2 - m_c{\rm Re}M'' \right) \notag \\
               &~~~+\frac{|M|^2}{4G}\bigg] + \mathcal{O}(m_c^2),
\end{align}
with the GL coefficients,
\begin{align}
 \alpha_{2j} =(-1)^j4N_cN_fT\sum_k \int\frac{d^3{\bf p}}{(2\pi)^3} \frac{1}{\left[(\omega_k + i\mu)^2 + {\bf p}^2\right]^j}.
\end{align}

Switching on $B$, the summation of the eigenstate and the energy spectrum in $\alpha_{2j}$ should change as the following,
\begin{align}
 &4N_cN_f \int \frac{d^3{\bf p}}{(2\pi)^3} \rightarrow 2N_c \sum_f \frac{|e_fB|}{2\pi} \sum_n \int \frac{dp_z}{2\pi}(2-\delta_{n,0}), \\
 &{\bf p}^2 \rightarrow p_z^2 + 2|e_fB|n,
\end{align}
where $n$ represents the Landau levels.
Furthermore some odd order terms are added.
The third order term is derived from a part of $j=2$ in Eq.~(\ref{ggl}),
\begin{align}
 &-\frac{T}{V}\frac{1}{2} {\rm{Tr}}_{D,c,f,V} \left[S_B\left( {\rm Re}\tilde{M} + i\gamma^5 \tau^3 {\rm Im}\tilde{M} \right) \right]^2 \notag \\
 \rightarrow& -\frac{T}{V}\frac{1}{2}N_c\sum_f \int d^4xd^4x' \notag \\
                & \times {\rm tr} \big\{ [{\rm Re}\tilde{M}(x_3) + i \gamma_5\sigma_f {\rm Im}\tilde{M}(x_3)]S_B(x,x')  \notag \\
                &~~~\times [{\rm Re}\tilde{M}'(x_3) + i \gamma_5\sigma_f {\rm Im}\tilde{M}'(x_3)](x'_3-x_3) S_B(x',x)  \big\} \notag \\
                =& N_c\sum_f\frac{|e_fB|}{16\pi^3 T} {\rm Im}\psi^{(1)}\left(\frac{1}{2}+i\frac{\mu}{2\pi T}\right)\int \frac{d^3\bm{x}}{V}{\rm Im}\left( \tilde{M}^*\tilde{M}' \right) \notag \\
                =& \tilde{\alpha}_3 \int \frac{d^3\bm{x}}{V} \left[ {\rm Im}\left( M^*M' \right) + m_c{\rm Im}M' \right],
\end{align}
with $\sigma_u=+1,\sigma_d=-1$.
It is convenient to use $S_B$ in the momentum representation \cite{schwinger,chodos}.
The fifth order terms are derived from a part of $j=2$ in Eq.~(\ref{ggl}),
\begin{align}
 &-\frac{T}{V}\frac{1}{2} {\rm{Tr}}_{D,c,f,V} \left[S_B\left( {\rm Re}\tilde{M} + i\gamma^5 \tau^3 {\rm Im}\tilde{M} \right) \right]^2 \notag \\
 \rightarrow& -\frac{T}{V}\frac{1}{2}N_c\sum_f \int d^4xd^4x' \notag \\
                & \times {\rm tr} \Big\{ [{\rm Re}\tilde{M}(x_3) + i \gamma_5\sigma_f {\rm Im}\tilde{M}(x_3)]S_B(x,x')  \notag \\
                &\times \frac{1}{6}[{\rm Re}\tilde{M}'''(x_3) + i \gamma_5\sigma_f {\rm Im}\tilde{M}'''(x_3)](x'_3-x_3) ^3 S_B(x',x) \Big\} \notag \\
                =& N_c\sum_f\frac{|e_fB|}{1536\pi^5 T^3} {\rm Im}\psi^{(3)}\left(\frac{1}{2}+i\frac{\mu}{2\pi T}\right) \int \frac{d^3\bm{x}}{V}{\rm Im}\left( \tilde{M}^*\tilde{M}''' \right) \notag \\
                \sim&\,\tilde{\alpha}_{4b} \int \frac{d^3\bm{x}}{V}{\rm Im}M'''.
\end{align}
From a part of $j=4$,
\begin{align}
 &-\frac{T}{V}\frac{1}{4} {\rm{Tr}}_{D,c,f,V} \left[S_B\left( {\rm Re}\tilde{M} + i\gamma^5 \tau^3 {\rm Im}\tilde{M} \right) \right]^4 \notag \\
 \rightarrow&-\frac{T}{\mathcal{V}}\frac{3}{4} N_c\sum_f \int d^4xd^4x'd^4x''d^4x''' \notag \\
                &\times{\rm tr} \Big\{  [{\rm Re}\tilde{M}(x_3) + i \gamma_5\sigma_f {\rm Im}\tilde{M}(x_3)]S_B(x,x') \notag \\
                &~~\times [{\rm Re}\tilde{M}(x_3) + i \gamma_5\sigma_f {\rm Im}\tilde{M}(x_3))] S_B(x',x'') \notag \\
                &~~\times [{\rm Re}\tilde{M}(x_3) + i \gamma_5\sigma_f {\rm Im}\tilde{M}(x_3)] S_B(x'',x''') \notag \\
                &~~\times [{\rm Re}\tilde{M}'(x_3) + i \gamma_5\sigma_f {\rm Im}\tilde{M}'(x_3)](x'''_3-x_3) S_B(x''',x) \Big\}. \notag \\
\end{align}
Here we can see that only $|\tilde{M}|^2{\rm Im}(\tilde{M}^*\tilde{M}')\sim m_c|M|^2{\rm Im}M' + 2m_c{\rm Re}M{\rm Im}(M^*M') $ term survives after taking the Dirac trace and integrating.
Therefore this term can be described as,
\begin{align}
 \tilde{\alpha}_{4a} \int \frac{d^3\bm{x}}{V}\left[ |M|^2{\rm Im}M' + 2{\rm Re}M{\rm Im}(M^*M')\right],
\end{align}
where the coefficient is written as $\tilde{\alpha}_{4a}$ for convenience. 
In summary, the thermodynamic potential to fourth order takes the Eq.~(\ref{omega}).

\section{Regularization of the GL coefficients}
In this appendix, the GL coefficients including divergence is regularized by PVR.
For convenience, we introduce the function $I_j$ and rewrite Eq.~(\ref{a2m}),
\begin{align}
 &\alpha_{2j} = (-1)^j N_c\sum_f \frac{|e_fB|}{2\pi} I_j(0),
\end{align}
where
\begin{align}
 &I_j(\Lambda^2) \equiv 2T\sum_{k}\sum_{n\geq0} \int \frac{dp}{2\pi}\frac{2-\delta_{n,0}}{\left[ (\omega_k + i\mu)^2 + E_n^2(\Lambda^2) \right]^j}, \\
 &E_n(\Lambda^2) \equiv \sqrt{p^2 + \Lambda^2 + 2|e_fB|n}.
\end{align}
Then, $I_1(0)$ and $I_2(0)$ should be regularized.
Taking the Matsubara summation,
\begin{align}
 I_1 &= \sum_{n\geq0} \int \frac{dp}{2\pi}\frac{2-\delta_{n,0}}{E_n} \left[ 1 - f_F(E_n+\mu) - f_F(E_n-\mu)  \right], \\
 I_2 &= \frac{1}{2} \sum_{n\geq0} (2-\delta_{n,0}) \notag \\
     & \times \int \frac{dp}{2\pi} \bigg\{ \frac{1}{E_n^3} \left[ 1 - f_F(E_n+\mu) - f_F(E_n-\mu) \right] \notag \\
     &~~~~~~~~~~~~~+ \frac{1}{E_n^2}\left[ f'_F(E_n+\mu) + f'_F(E_n-\mu) \right] \bigg\},
\end{align}
where $f_F$ is the Fermion distribution function.
Therefore the diverging vacuum part can be decomposed into the form,
\begin{align}
 I_{1,{\rm vac}} &= \sum_{n\geq0} \int \frac{dp}{2\pi}\frac{2-\delta_{n,0}}{E_n}, \\
 I_{2,{\rm vac}} &= \frac{1}{2} \sum_{n\geq0} \int \frac{dp}{2\pi}\frac{2-\delta_{n,0}}{E_n^3}.
\end{align}
Then, $I_1(0)$ and $I_2(0)$ are regularized as the following,
\begin{align}
 I_{1,{\rm vac}}(0) &\rightarrow I_{1,{\rm vac}}(0) - 2I_{1,{\rm vac}}(\Lambda^2) + I_{1,{\rm vac}}(2\Lambda^2), \\
 I_{2,{\rm vac}}(0) &\rightarrow I_{2,{\rm vac}}(0) - I_{2,{\rm vac}}(\Lambda^2).
\end{align}
Thus, all divergence of coefficients can be excluded. 

\section{Spectral asymmetry with $m_c$}
In this appendix, we show that $\tilde{\alpha}_3$ term is derived from the spectral asymmetry and relevant to the chiral anomaly when the inhomogeneous chiral condensate has the degree of freedom of the phase.
Generally quark number with the finite $T$ is given as \cite{tatsumi},
\begin{align}
 N = -\frac{1}{2}\eta_H + \int dE \rho(E) \left[ \frac{\theta(E)}{1+e^{\beta(E-\mu)}} - \frac{\theta(-E)}{1+e^{-\beta(E+\mu)}} \right],
\end{align}
where $\rho(E)$ is the density of state.
The first term, which is called the Atiyah-Patodi-Singer $\eta$-invariant represents the anomalous particle number \cite{niemi,niemi2},
\begin{align}
 \eta_H = \lim_{s\to +0}\int dE \rho(E) {\rm sign}(E)|E|^{-s},
\end{align}
and measures the extent of {\it spectral asymmetry} about zero.
The second term $(N_{\rm nom})$ corresponds to the normal particle number and we rewrite it as the form including the summation of the Matsubara frequency,
\begin{align}
 N_{\rm nom} = \frac{1}{2} \eta_H - \int dE \rho(E) T\sum_k \frac{1}{E-\mu-i\omega_k}. \label{nom}
\end{align}
Here we can see that the first term in (\ref{nom}) cancels out the anomalous particle number.
However, the information of the $\eta$ invariant is not washed away
since the infinite series reproduces the anomalous particle number at $\mu=T=0$.

The local density of state takes the form,
\begin{align}
 \rho({\bf x},E) &= \frac{1}{\pi} {\rm Im}\,{\rm tr}_{D,f,c} [R({\bf x},E+i\epsilon)] \notag \\
                     &= -\frac{N_c}{\pi}\sum_f \frac{\partial}{\partial E}{\rm Im}\,{\rm tr}_D \left\langle {\bf x}\left|\ln(H-E-i\epsilon)\right|{\bf x}\right\rangle,
\end{align}
with the resolvent:~$R({\bf x},E)\equiv \left\langle{\bf x}\left|\frac{1}{H-E}\right|{\bf x}\right\rangle$.
In the present model, Hamiltonian takes the form,
\begin{align}
 H = \vec{\alpha} \cdot {\bf P} + \gamma^0 \left[ m_c + m e^{i\gamma^5\tau_3\theta({\bf r})} \right],
\end{align}
where $\alpha_i = \gamma_0\gamma_i$ and ${\bf P}$ is the covariant derivative.
After the Weinberg transformation:~$\psi \rightarrow \psi_W=e^{i\gamma^5\tau_3\theta({\bf r})/2}\psi$, Hamiltonian changes to $\tilde{H}$,
\begin{align}
 \tilde{H} &= \tilde{H}_0 + \delta \tilde{H}, \\
 \tilde{H}_0 &\equiv \vec{\alpha} \cdot {\bf P} + \gamma_0 m, \\
 \delta\tilde{H} &\equiv \gamma^0\left[ m_c e^{-i\gamma^5\tau_3\theta({\bf r})} - \frac{1}{2}\gamma^5\tau_3 \vec{\gamma} \cdot {\bf \nabla} \theta({\bf r}) \right].
\end{align}
Therefore, $\rho(\bm{x},E)$ can be expanded to the form, 
\begin{align}
 \rho({\bf x},E) &= \frac{N_c}{\pi} \sum_f {\rm Im}\,{\rm tr} \left\langle {\bf x}\left| \frac{1}{\tilde{H}_0-E} \right|{\bf x}\right\rangle \notag \\
                     &~- \frac{N_c}{\pi} \sum_f \frac{\partial}{\partial E} {\rm Im}\,{\rm tr} \left\langle {\bf x}\left| \frac{1}{\tilde{H}_0-E-i\epsilon} \right|{\bf x}\right\rangle \delta\tilde{H}({\bf x}) \notag \\
                     &~ + {\cal O}\left( \partial(\delta\tilde{H}), (\delta\tilde{H})^2 \right),
\end{align}
where the first term does not depend on $\theta$.
Here, $\left\langle {\bf x}\left| \frac{1}{\tilde{H}_0-E-i\epsilon} \right|{\bf x}\right\rangle$ can be rewritten into the propagator decomposed over the Landau levels \cite{tatsumi}.

Then the reading term proportional to $\partial\theta$ takes the form,
\begin{align}
 &\rho_{\partial\theta}({\bf x},E) \notag \\
 &~~~= -\frac{N_c}{4\pi^2}\sum_f |e_fB|\partial_z\theta({\bf x})\frac{\partial}{\partial E} \left[ \frac{ |E|}{\sqrt{E^2 - m^2}}\theta\left(|E|-m\right) \right].
\end{align}
From the Eq.~(\ref{nom}), the part of quark number generated by $\rho_{\partial \theta}$ takes the form,
\begin{align}
 &N_{\partial\theta} = \frac{N_c}{4\pi^2}\sum_f |e_fB|\int d^3{\bf x}\partial_z\theta({\bf x}) \Bigg\{1 + T\sum_k \int_0^\infty dy  \notag \\
 &\times \left[ \frac{1}{(\sqrt{y^2+m^2}-\mu-i\omega_k)^2} + \frac{1}{(\sqrt{y^2+m^2}+\mu+i\omega_k)^2} \right] \Bigg\}, \label{numdel}
\end{align}
where the first term is derived from the surface term in the partial integral about $E$
and we take $y=\sqrt{E^2-m^2}$.
Then the second term can be expanded with respect to $m^2$.
It can be seen that $m^0$ part of the second term cancels out the first term and the the remnant of $N_{\partial\theta}$ takes the form,
\begin{align}
 N_{\partial\theta} =& -\frac{N_c}{16\pi^3T}\sum_f|e_fB| \int d{\bf x}^3\partial_z\theta({\bf x}) \notag \\
                           &\times \frac{\partial}{\partial \mu}{\rm Im} \psi^{(1)}\left(\frac{1}{2} + i\frac{\mu}{2\pi T} \right) m^2 + \mathcal{O}(m^4).
\end{align}
From the thermodynamic relation: $N/V = -\partial \Omega / \partial \mu$,
we can see that $\tilde{\alpha}_3$ term is generated.

On the other hand, the result from the chiral anomaly \cite{son} is recovered in the limit: $m\rightarrow \infty$.
Then the second term in the Eq.~(\ref{numdel}) vanishes and the first term is the very contribution of chiral anomaly.
This limit is consistent with the case where there is no valence quarks argued in the Ref.~\cite{yoshiike}.
Furthermore, substituting the configuration of $\theta$ (\ref{sol}), the quark number takes the form,
\begin{align}
 N_{\partial\theta} &\rightarrow \frac{N_c}{4\pi^2}\sum_f |e_fB| \frac{\pi m^*_\pi}{kK(k)}.
\end{align}
For investigating the variation from the case of chiral limit \cite{tatsumi},
we take $2m^*_\pi/k=q$, where $q$ is the wave vector of the DCDW condensate.
Then it can be expanded with respect to $\left(m_\pi^*/q\right)^2$,
\begin{align}
 N_{\partial\theta} &= \frac{N_c}{4\pi^2}\sum_f |e_fB|q \left[ 1 - 2\frac{m_\pi^{*2}}{q^2} + \mathcal{O}\left( \frac{m_\pi^{*4}}{q^4} \right)\right].
\end{align}
The second term represents the correction by the finite $m_c$ because of $m_\pi^{*2}\sim m_c$.
The result also implies that the spectral asymmetry has the correction $\mathcal{O}(m_c)$
although the exact energy spectrum cannot be obtained at the finite $m_c$.

\bibliographystyle{apsrev4-1}
\bibliography{reference}

\begin{thebibliography}{53}%
\makeatletter
\providecommand \@ifxundefined [1]{%
 \@ifx{#1\undefined}
}%
\providecommand \@ifnum [1]{%
 \ifnum #1\expandafter \@firstoftwo
 \else \expandafter \@secondoftwo
 \fi
}%
\providecommand \@ifx [1]{%
 \ifx #1\expandafter \@firstoftwo
 \else \expandafter \@secondoftwo
 \fi
}%
\providecommand \natexlab [1]{#1}%
\providecommand \enquote  [1]{``#1''}%
\providecommand \bibnamefont  [1]{#1}%
\providecommand \bibfnamefont [1]{#1}%
\providecommand \citenamefont [1]{#1}%
\providecommand \href@noop [0]{\@secondoftwo}%
\providecommand \href [0]{\begingroup \@sanitize@url \@href}%
\providecommand \@href[1]{\@@startlink{#1}\@@href}%
\providecommand \@@href[1]{\endgroup#1\@@endlink}%
\providecommand \@sanitize@url [0]{\catcode `\\12\catcode `\$12\catcode
  `\&12\catcode `\#12\catcode `\^12\catcode `\_12\catcode `\%12\relax}%
\providecommand \@@startlink[1]{}%
\providecommand \@@endlink[0]{}%
\providecommand \url  [0]{\begingroup\@sanitize@url \@url }%
\providecommand \@url [1]{\endgroup\@href {#1}{\urlprefix }}%
\providecommand \urlprefix  [0]{URL }%
\providecommand \Eprint [0]{\href }%
\providecommand \doibase [0]{http://dx.doi.org/}%
\providecommand \selectlanguage [0]{\@gobble}%
\providecommand \bibinfo  [0]{\@secondoftwo}%
\providecommand \bibfield  [0]{\@secondoftwo}%
\providecommand \translation [1]{[#1]}%
\providecommand \BibitemOpen [0]{}%
\providecommand \bibitemStop [0]{}%
\providecommand \bibitemNoStop [0]{.\EOS\space}%
\providecommand \EOS [0]{\spacefactor3000\relax}%
\providecommand \BibitemShut  [1]{\csname bibitem#1\endcsname}%
\let\auto@bib@innerbib\@empty
\bibitem [{\citenamefont {Nakano}\ and\ \citenamefont
  {Tatsumi}(2005)}]{nakano}%
  \BibitemOpen
  \bibfield  {author} {\bibinfo {author} {\bibfnamefont {E.}~\bibnamefont
  {Nakano}}\ and\ \bibinfo {author} {\bibfnamefont {T.}~\bibnamefont
  {Tatsumi}},\ }\href@noop {} {\bibfield  {journal} {\bibinfo  {journal} {Phys.
  Rev. D}\ }\textbf {\bibinfo {volume} {71}},\ \bibinfo {pages} {114006}
  (\bibinfo {year} {2005})}\BibitemShut {NoStop}%
\bibitem [{\citenamefont {Nickel}(2009{\natexlab{a}})}]{nickel}%
  \BibitemOpen
  \bibfield  {author} {\bibinfo {author} {\bibfnamefont {D.}~\bibnamefont
  {Nickel}},\ }\href@noop {} {\bibfield  {journal} {\bibinfo  {journal} {Phys.
  Rev. D}\ }\textbf {\bibinfo {volume} {80}},\ \bibinfo {pages} {074025}
  (\bibinfo {year} {2009}{\natexlab{a}})}\BibitemShut {NoStop}%
\bibitem [{\citenamefont {Nickel}(2009{\natexlab{b}})}]{ggl}%
  \BibitemOpen
  \bibfield  {author} {\bibinfo {author} {\bibfnamefont {D.}~\bibnamefont
  {Nickel}},\ }\href {\doibase 10.1103/PhysRevLett.103.072301} {\bibfield
  {journal} {\bibinfo  {journal} {Phys. Rev. Lett.}\ }\textbf {\bibinfo
  {volume} {103}},\ \bibinfo {pages} {072301} (\bibinfo {year}
  {2009}{\natexlab{b}})}\BibitemShut {NoStop}%
\bibitem [{\citenamefont {Muller}\ \emph {et~al.}(2013)\citenamefont {Muller},
  \citenamefont {Buballa},\ and\ \citenamefont {Wambach}}]{muller}%
  \BibitemOpen
  \bibfield  {author} {\bibinfo {author} {\bibfnamefont {D.}~\bibnamefont
  {Muller}}, \bibinfo {author} {\bibfnamefont {M.}~\bibnamefont {Buballa}}, \
  and\ \bibinfo {author} {\bibfnamefont {J.}~\bibnamefont {Wambach}},\
  }\href@noop {} {\bibfield  {journal} {\bibinfo  {journal} {Phys. Lett. B}\
  }\textbf {\bibinfo {volume} {727}},\ \bibinfo {pages} {240} (\bibinfo {year}
  {2013})}\BibitemShut {NoStop}%
\bibitem [{\citenamefont {Nishiyama}\ \emph {et~al.}(2015)\citenamefont
  {Nishiyama}, \citenamefont {Karasawa},\ and\ \citenamefont
  {Tatsumi}}]{nishiyama}%
  \BibitemOpen
  \bibfield  {author} {\bibinfo {author} {\bibfnamefont {K.}~\bibnamefont
  {Nishiyama}}, \bibinfo {author} {\bibfnamefont {S.}~\bibnamefont {Karasawa}},
  \ and\ \bibinfo {author} {\bibfnamefont {T.}~\bibnamefont {Tatsumi}},\ }\href
  {\doibase 10.1103/PhysRevD.92.036008} {\bibfield  {journal} {\bibinfo
  {journal} {Phys. Rev. D}\ }\textbf {\bibinfo {volume} {92}},\ \bibinfo
  {pages} {036008} (\bibinfo {year} {2015})}\BibitemShut {NoStop}%
\bibitem [{\citenamefont {Basar}\ and\ \citenamefont {Dunne}(2008)}]{basar}%
  \BibitemOpen
  \bibfield  {author} {\bibinfo {author} {\bibfnamefont {G.}~\bibnamefont
  {Basar}}\ and\ \bibinfo {author} {\bibfnamefont {G.~V.}\ \bibnamefont
  {Dunne}},\ }\href@noop {} {\bibfield  {journal} {\bibinfo  {journal} {Phys.
  Rev. D}\ }\textbf {\bibinfo {volume} {78}},\ \bibinfo {pages} {065022}
  (\bibinfo {year} {2008})}\BibitemShut {NoStop}%
\bibitem [{\citenamefont {Basar}\ \emph {et~al.}(2009)\citenamefont {Basar},
  \citenamefont {Dunne},\ and\ \citenamefont {Thies}}]{basar2}%
  \BibitemOpen
  \bibfield  {author} {\bibinfo {author} {\bibfnamefont {G.}~\bibnamefont
  {Basar}}, \bibinfo {author} {\bibfnamefont {G.~V.}\ \bibnamefont {Dunne}}, \
  and\ \bibinfo {author} {\bibfnamefont {M.}~\bibnamefont {Thies}},\
  }\href@noop {} {\bibfield  {journal} {\bibinfo  {journal} {Phys. Rev. D}\
  }\textbf {\bibinfo {volume} {79}},\ \bibinfo {pages} {105012} (\bibinfo
  {year} {2009})}\BibitemShut {NoStop}%
\bibitem [{\citenamefont {Suganuma}\ and\ \citenamefont
  {Tatsumi}(1991)}]{suganuma}%
  \BibitemOpen
  \bibfield  {author} {\bibinfo {author} {\bibfnamefont {H.}~\bibnamefont
  {Suganuma}}\ and\ \bibinfo {author} {\bibfnamefont {T.}~\bibnamefont
  {Tatsumi}},\ }\href@noop {} {\bibfield  {journal} {\bibinfo  {journal} {Ann.
  Phys.}\ }\textbf {\bibinfo {volume} {208}},\ \bibinfo {pages} {470} (\bibinfo
  {year} {1991})}\BibitemShut {NoStop}%
\bibitem [{\citenamefont {Klevansky}\ and\ \citenamefont
  {Lemmer}(1989)}]{klevansky}%
  \BibitemOpen
  \bibfield  {author} {\bibinfo {author} {\bibfnamefont {S.~P.}\ \bibnamefont
  {Klevansky}}\ and\ \bibinfo {author} {\bibfnamefont {R.~H.}\ \bibnamefont
  {Lemmer}},\ }\href {\doibase 10.1103/PhysRevD.39.3478} {\bibfield  {journal}
  {\bibinfo  {journal} {Phys. Rev. D}\ }\textbf {\bibinfo {volume} {39}},\
  \bibinfo {pages} {3478} (\bibinfo {year} {1989})}\BibitemShut {NoStop}%
\bibitem [{\citenamefont {Gusynin}\ \emph {et~al.}(1996)\citenamefont
  {Gusynin}, \citenamefont {Miransky},\ and\ \citenamefont
  {Shovkovy}}]{gusynin}%
  \BibitemOpen
  \bibfield  {author} {\bibinfo {author} {\bibfnamefont {V.}~\bibnamefont
  {Gusynin}}, \bibinfo {author} {\bibfnamefont {V.}~\bibnamefont {Miransky}}, \
  and\ \bibinfo {author} {\bibfnamefont {I.}~\bibnamefont {Shovkovy}},\ }\href
  {\doibase http://dx.doi.org/10.1016/0550-3213(96)00021-1} {\bibfield
  {journal} {\bibinfo  {journal} {Nucl. Phys. B}\ }\textbf {\bibinfo {volume}
  {462}},\ \bibinfo {pages} {249 } (\bibinfo {year} {1996})}\BibitemShut
  {NoStop}%
\bibitem [{\citenamefont {Bali}\ \emph {et~al.}(2012)\citenamefont {Bali},
  \citenamefont {Bruckmann}, \citenamefont {Endr\ifmmode~\mbox{\H{o}}\else
  \H{o}\fi{}di}, \citenamefont {Fodor}, \citenamefont {Katz},\ and\
  \citenamefont {Sch\"afer}}]{bali}%
  \BibitemOpen
  \bibfield  {author} {\bibinfo {author} {\bibfnamefont {G.~S.}\ \bibnamefont
  {Bali}}, \bibinfo {author} {\bibfnamefont {F.}~\bibnamefont {Bruckmann}},
  \bibinfo {author} {\bibfnamefont {G.}~\bibnamefont
  {Endr\ifmmode~\mbox{\H{o}}\else \H{o}\fi{}di}}, \bibinfo {author}
  {\bibfnamefont {Z.}~\bibnamefont {Fodor}}, \bibinfo {author} {\bibfnamefont
  {S.~D.}\ \bibnamefont {Katz}}, \ and\ \bibinfo {author} {\bibfnamefont
  {A.}~\bibnamefont {Sch\"afer}},\ }\href {\doibase 10.1103/PhysRevD.86.071502}
  {\bibfield  {journal} {\bibinfo  {journal} {Phys. Rev. D}\ }\textbf {\bibinfo
  {volume} {86}},\ \bibinfo {pages} {071502} (\bibinfo {year}
  {2012})}\BibitemShut {NoStop}%
\bibitem [{\citenamefont {Endrodi}(2015)}]{endrodi}%
  \BibitemOpen
  \bibfield  {author} {\bibinfo {author} {\bibfnamefont {G.}~\bibnamefont
  {Endrodi}},\ }\href@noop {} {\bibfield  {journal} {\bibinfo  {journal}
  {JHEP}\ }\textbf {\bibinfo {volume} {1507}},\ \bibinfo {eid} {173} (\bibinfo
  {year} {2015})}\BibitemShut {NoStop}%
\bibitem [{\citenamefont {Farias}\ \emph {et~al.}(2014)\citenamefont {Farias},
  \citenamefont {Gomes}, \citenamefont {Krein},\ and\ \citenamefont
  {Pinto}}]{farias}%
  \BibitemOpen
  \bibfield  {author} {\bibinfo {author} {\bibfnamefont {R.~L.~S.}\
  \bibnamefont {Farias}}, \bibinfo {author} {\bibfnamefont {K.~P.}\
  \bibnamefont {Gomes}}, \bibinfo {author} {\bibfnamefont {G.}~\bibnamefont
  {Krein}}, \ and\ \bibinfo {author} {\bibfnamefont {M.~B.}\ \bibnamefont
  {Pinto}},\ }\href {\doibase 10.1103/PhysRevC.90.025203} {\bibfield  {journal}
  {\bibinfo  {journal} {Phys. Rev. C}\ }\textbf {\bibinfo {volume} {90}},\
  \bibinfo {pages} {025203} (\bibinfo {year} {2014})}\BibitemShut {NoStop}%
\bibitem [{\citenamefont {Ferrer}\ \emph {et~al.}(2015)\citenamefont {Ferrer},
  \citenamefont {de~la Incera},\ and\ \citenamefont {Wen}}]{ferrer}%
  \BibitemOpen
  \bibfield  {author} {\bibinfo {author} {\bibfnamefont {E.~J.}\ \bibnamefont
  {Ferrer}}, \bibinfo {author} {\bibfnamefont {V.}~\bibnamefont {de~la
  Incera}}, \ and\ \bibinfo {author} {\bibfnamefont {X.~J.}\ \bibnamefont
  {Wen}},\ }\href {\doibase 10.1103/PhysRevD.91.054006} {\bibfield  {journal}
  {\bibinfo  {journal} {Phys. Rev. D}\ }\textbf {\bibinfo {volume} {91}},\
  \bibinfo {pages} {054006} (\bibinfo {year} {2015})}\BibitemShut {NoStop}%
\bibitem [{\citenamefont {Mueller}\ and\ \citenamefont
  {Pawlowski}(2015)}]{mueller}%
  \BibitemOpen
  \bibfield  {author} {\bibinfo {author} {\bibfnamefont {N.}~\bibnamefont
  {Mueller}}\ and\ \bibinfo {author} {\bibfnamefont {J.~M.}\ \bibnamefont
  {Pawlowski}},\ }\href {\doibase 10.1103/PhysRevD.91.116010} {\bibfield
  {journal} {\bibinfo  {journal} {Phys. Rev. D}\ }\textbf {\bibinfo {volume}
  {91}},\ \bibinfo {pages} {116010} (\bibinfo {year} {2015})}\BibitemShut
  {NoStop}%
\bibitem [{\citenamefont {Braun}\ \emph {et~al.}(2014)\citenamefont {Braun},
  \citenamefont {Mian},\ and\ \citenamefont {Rechenberger}}]{braun}%
  \BibitemOpen
  \bibfield  {author} {\bibinfo {author} {\bibfnamefont {J.}~\bibnamefont
  {Braun}}, \bibinfo {author} {\bibfnamefont {W.~A.}\ \bibnamefont {Mian}}, \
  and\ \bibinfo {author} {\bibfnamefont {S.}~\bibnamefont {Rechenberger}},\
  }\href@noop {} {\bibfield  {journal} {\bibinfo  {journal} {arXiv:1412.6025}\
  } (\bibinfo {year} {2014})}\BibitemShut {NoStop}%
\bibitem [{\citenamefont {Frolov}\ \emph {et~al.}(2010)\citenamefont {Frolov},
  \citenamefont {Zhukovsky},\ and\ \citenamefont {Klimenko}}]{frolov}%
  \BibitemOpen
  \bibfield  {author} {\bibinfo {author} {\bibfnamefont {I.~E.}\ \bibnamefont
  {Frolov}}, \bibinfo {author} {\bibfnamefont {V.~C.}\ \bibnamefont
  {Zhukovsky}}, \ and\ \bibinfo {author} {\bibfnamefont {K.}~\bibnamefont
  {Klimenko}},\ }\href@noop {} {\bibfield  {journal} {\bibinfo  {journal}
  {Phys. Rev. D}\ }\textbf {\bibinfo {volume} {82}},\ \bibinfo {pages} {076002}
  (\bibinfo {year} {2010})}\BibitemShut {NoStop}%
\bibitem [{\citenamefont {Yoshiike}\ \emph {et~al.}(2015)\citenamefont
  {Yoshiike}, \citenamefont {Nishiyama},\ and\ \citenamefont
  {Tatsumi}}]{yoshiike}%
  \BibitemOpen
  \bibfield  {author} {\bibinfo {author} {\bibfnamefont {R.}~\bibnamefont
  {Yoshiike}}, \bibinfo {author} {\bibfnamefont {K.}~\bibnamefont {Nishiyama}},
  \ and\ \bibinfo {author} {\bibfnamefont {T.}~\bibnamefont {Tatsumi}},\ }\href
  {\doibase http://dx.doi.org/10.1016/j.physletb.2015.10.028} {\bibfield
  {journal} {\bibinfo  {journal} {Phys. Lett. B}\ }\textbf {\bibinfo {volume}
  {751}},\ \bibinfo {pages} {123 } (\bibinfo {year} {2015})},\ \Eprint
  {http://arxiv.org/abs/1507.02110} {arXiv:1507.02110 [hep-ph]} \BibitemShut
  {NoStop}%
\bibitem [{\citenamefont {Tatsumi}\ \emph {et~al.}(2015)\citenamefont
  {Tatsumi}, \citenamefont {Nishiyama},\ and\ \citenamefont
  {Karasawa}}]{tatsumi}%
  \BibitemOpen
  \bibfield  {author} {\bibinfo {author} {\bibfnamefont {T.}~\bibnamefont
  {Tatsumi}}, \bibinfo {author} {\bibfnamefont {K.}~\bibnamefont {Nishiyama}},
  \ and\ \bibinfo {author} {\bibfnamefont {S.}~\bibnamefont {Karasawa}},\
  }\href {\doibase http://dx.doi.org/10.1016/j.physletb.2015.02.033} {\bibfield
   {journal} {\bibinfo  {journal} {Phys. Lett. B}\ }\textbf {\bibinfo {volume}
  {743}},\ \bibinfo {pages} {66 } (\bibinfo {year} {2015})}\BibitemShut
  {NoStop}%
\bibitem [{\citenamefont {Allton}\ \emph {et~al.}(2005)\citenamefont {Allton},
  \citenamefont {Doring}, \citenamefont {Ejiri}, \citenamefont {Hands},
  \citenamefont {Kaczmarek} \emph {et~al.}}]{allton}%
  \BibitemOpen
  \bibfield  {author} {\bibinfo {author} {\bibfnamefont {C.}~\bibnamefont
  {Allton}}, \bibinfo {author} {\bibfnamefont {M.}~\bibnamefont {Doring}},
  \bibinfo {author} {\bibfnamefont {S.}~\bibnamefont {Ejiri}}, \bibinfo
  {author} {\bibfnamefont {S.}~\bibnamefont {Hands}}, \bibinfo {author}
  {\bibfnamefont {O.}~\bibnamefont {Kaczmarek}},  \emph {et~al.},\ }\href
  {\doibase 10.1103/PhysRevD.71.054508} {\bibfield  {journal} {\bibinfo
  {journal} {Phys.Rev.}\ }\textbf {\bibinfo {volume} {D71}},\ \bibinfo {pages}
  {054508} (\bibinfo {year} {2005})},\ \Eprint
  {http://arxiv.org/abs/hep-lat/0501030} {arXiv:hep-lat/0501030 [hep-lat]}
  \BibitemShut {NoStop}%
\bibitem [{\citenamefont {Gavai}\ and\ \citenamefont {Gupta}(2008)}]{gavai}%
  \BibitemOpen
  \bibfield  {author} {\bibinfo {author} {\bibfnamefont {R.}~\bibnamefont
  {Gavai}}\ and\ \bibinfo {author} {\bibfnamefont {S.}~\bibnamefont {Gupta}},\
  }\href {\doibase 10.1103/PhysRevD.78.114503} {\bibfield  {journal} {\bibinfo
  {journal} {Phys.Rev.}\ }\textbf {\bibinfo {volume} {D78}},\ \bibinfo {pages}
  {114503} (\bibinfo {year} {2008})},\ \Eprint {http://arxiv.org/abs/0806.2233}
  {arXiv:0806.2233 [hep-lat]} \BibitemShut {NoStop}%
\bibitem [{\citenamefont {Fodor}\ and\ \citenamefont
  {Katz}(2002{\natexlab{a}})}]{fodor1}%
  \BibitemOpen
  \bibfield  {author} {\bibinfo {author} {\bibfnamefont {Z.}~\bibnamefont
  {Fodor}}\ and\ \bibinfo {author} {\bibfnamefont {S.}~\bibnamefont {Katz}},\
  }\href {\doibase 10.1016/S0370-2693(02)01583-6} {\bibfield  {journal}
  {\bibinfo  {journal} {Phys.Lett.}\ }\textbf {\bibinfo {volume} {B534}},\
  \bibinfo {pages} {87} (\bibinfo {year} {2002}{\natexlab{a}})},\ \Eprint
  {http://arxiv.org/abs/hep-lat/0104001} {arXiv:hep-lat/0104001 [hep-lat]}
  \BibitemShut {NoStop}%
\bibitem [{\citenamefont {Fodor}\ \emph {et~al.}(2003)\citenamefont {Fodor},
  \citenamefont {Katz},\ and\ \citenamefont {Szabo}}]{fodor2}%
  \BibitemOpen
  \bibfield  {author} {\bibinfo {author} {\bibfnamefont {Z.}~\bibnamefont
  {Fodor}}, \bibinfo {author} {\bibfnamefont {S.}~\bibnamefont {Katz}}, \ and\
  \bibinfo {author} {\bibfnamefont {K.}~\bibnamefont {Szabo}},\ }\href
  {\doibase 10.1016/j.physletb.2003.06.011} {\bibfield  {journal} {\bibinfo
  {journal} {Phys.Lett.}\ }\textbf {\bibinfo {volume} {B568}},\ \bibinfo
  {pages} {73} (\bibinfo {year} {2003})},\ \Eprint
  {http://arxiv.org/abs/hep-lat/0208078} {arXiv:hep-lat/0208078 [hep-lat]}
  \BibitemShut {NoStop}%
\bibitem [{\citenamefont {Fodor}\ and\ \citenamefont
  {Katz}(2002{\natexlab{b}})}]{fodor3}%
  \BibitemOpen
  \bibfield  {author} {\bibinfo {author} {\bibfnamefont {Z.}~\bibnamefont
  {Fodor}}\ and\ \bibinfo {author} {\bibfnamefont {S.}~\bibnamefont {Katz}},\
  }\href {\doibase 10.1088/1126-6708/2002/03/014} {\bibfield  {journal}
  {\bibinfo  {journal} {JHEP}\ }\textbf {\bibinfo {volume} {0203}},\ \bibinfo
  {pages} {014} (\bibinfo {year} {2002}{\natexlab{b}})},\ \Eprint
  {http://arxiv.org/abs/hep-lat/0106002} {arXiv:hep-lat/0106002 [hep-lat]}
  \BibitemShut {NoStop}%
\bibitem [{\citenamefont {Fodor}\ and\ \citenamefont {Katz}(2004)}]{fodor4}%
  \BibitemOpen
  \bibfield  {author} {\bibinfo {author} {\bibfnamefont {Z.}~\bibnamefont
  {Fodor}}\ and\ \bibinfo {author} {\bibfnamefont {S.}~\bibnamefont {Katz}},\
  }\href {\doibase 10.1088/1126-6708/2004/04/050} {\bibfield  {journal}
  {\bibinfo  {journal} {JHEP}\ }\textbf {\bibinfo {volume} {0404}},\ \bibinfo
  {pages} {050} (\bibinfo {year} {2004})},\ \Eprint
  {http://arxiv.org/abs/hep-lat/0402006} {arXiv:hep-lat/0402006 [hep-lat]}
  \BibitemShut {NoStop}%
\bibitem [{\citenamefont {Alexandru}\ \emph {et~al.}(2005)\citenamefont
  {Alexandru}, \citenamefont {Faber}, \citenamefont {Horvath},\ and\
  \citenamefont {Liu}}]{alexandru}%
  \BibitemOpen
  \bibfield  {author} {\bibinfo {author} {\bibfnamefont {A.}~\bibnamefont
  {Alexandru}}, \bibinfo {author} {\bibfnamefont {M.}~\bibnamefont {Faber}},
  \bibinfo {author} {\bibfnamefont {I.}~\bibnamefont {Horvath}}, \ and\
  \bibinfo {author} {\bibfnamefont {K.-F.}\ \bibnamefont {Liu}},\ }\href
  {\doibase 10.1103/PhysRevD.72.114513} {\bibfield  {journal} {\bibinfo
  {journal} {Phys.Rev.}\ }\textbf {\bibinfo {volume} {D72}},\ \bibinfo {pages}
  {114513} (\bibinfo {year} {2005})},\ \Eprint
  {http://arxiv.org/abs/hep-lat/0507020} {arXiv:hep-lat/0507020 [hep-lat]}
  \BibitemShut {NoStop}%
\bibitem [{\citenamefont {Kratochvila}\ and\ \citenamefont
  {de~Forcrand}(2006)}]{kratochvila}%
  \BibitemOpen
  \bibfield  {author} {\bibinfo {author} {\bibfnamefont {S.}~\bibnamefont
  {Kratochvila}}\ and\ \bibinfo {author} {\bibfnamefont {P.}~\bibnamefont
  {de~Forcrand}},\ }\href {\doibase 10.1103/PhysRevD.73.114512} {\bibfield
  {journal} {\bibinfo  {journal} {Phys.Rev.}\ }\textbf {\bibinfo {volume}
  {D73}},\ \bibinfo {pages} {114512} (\bibinfo {year} {2006})},\ \Eprint
  {http://arxiv.org/abs/hep-lat/0602005} {arXiv:hep-lat/0602005 [hep-lat]}
  \BibitemShut {NoStop}%
\bibitem [{\citenamefont {de~Forcrand}\ and\ \citenamefont
  {Kratochvila}(2006)}]{forcrand3}%
  \BibitemOpen
  \bibfield  {author} {\bibinfo {author} {\bibfnamefont {P.}~\bibnamefont
  {de~Forcrand}}\ and\ \bibinfo {author} {\bibfnamefont {S.}~\bibnamefont
  {Kratochvila}},\ }\href {\doibase 10.1016/j.nuclphysbps.2006.01.007}
  {\bibfield  {journal} {\bibinfo  {journal} {Nucl.Phys.Proc.Suppl.}\ }\textbf
  {\bibinfo {volume} {153}},\ \bibinfo {pages} {62} (\bibinfo {year} {2006})},\
  \Eprint {http://arxiv.org/abs/hep-lat/0602024} {arXiv:hep-lat/0602024
  [hep-lat]} \BibitemShut {NoStop}%
\bibitem [{\citenamefont {Li}\ \emph {et~al.}(2010)\citenamefont {Li},
  \citenamefont {Alexandru}, \citenamefont {Liu},\ and\ \citenamefont
  {Meng}}]{li}%
  \BibitemOpen
  \bibfield  {author} {\bibinfo {author} {\bibfnamefont {A.}~\bibnamefont
  {Li}}, \bibinfo {author} {\bibfnamefont {A.}~\bibnamefont {Alexandru}},
  \bibinfo {author} {\bibfnamefont {K.-F.}\ \bibnamefont {Liu}}, \ and\
  \bibinfo {author} {\bibfnamefont {X.}~\bibnamefont {Meng}},\ }\href {\doibase
  10.1103/PhysRevD.82.054502} {\bibfield  {journal} {\bibinfo  {journal}
  {Phys.Rev.}\ }\textbf {\bibinfo {volume} {D82}},\ \bibinfo {pages} {054502}
  (\bibinfo {year} {2010})},\ \Eprint {http://arxiv.org/abs/1005.4158}
  {arXiv:1005.4158 [hep-lat]} \BibitemShut {NoStop}%
\bibitem [{\citenamefont {Barbour}\ and\ \citenamefont {Bell}(1992)}]{barbour}%
  \BibitemOpen
  \bibfield  {author} {\bibinfo {author} {\bibfnamefont {I.}~\bibnamefont
  {Barbour}}\ and\ \bibinfo {author} {\bibfnamefont {A.}~\bibnamefont {Bell}},\
  }\href {\doibase 10.1016/0550-3213(92)90324-5} {\bibfield  {journal}
  {\bibinfo  {journal} {Nucl.Phys.}\ }\textbf {\bibinfo {volume} {B372}},\
  \bibinfo {pages} {385} (\bibinfo {year} {1992})}\BibitemShut {NoStop}%
\bibitem [{\citenamefont {Nakamura}\ and\ \citenamefont
  {Nagata}(2013)}]{nakamura}%
  \BibitemOpen
  \bibfield  {author} {\bibinfo {author} {\bibfnamefont {A.}~\bibnamefont
  {Nakamura}}\ and\ \bibinfo {author} {\bibfnamefont {K.}~\bibnamefont
  {Nagata}},\ }\href@noop {} {\  (\bibinfo {year} {2013})},\ \Eprint
  {http://arxiv.org/abs/1305.0760} {arXiv:1305.0760 [hep-ph]} \BibitemShut
  {NoStop}%
\bibitem [{\citenamefont {Nagata}\ \emph {et~al.}(2015)\citenamefont {Nagata},
  \citenamefont {Kashiwa}, \citenamefont {Nakamura},\ and\ \citenamefont
  {Nishigaki}}]{nagata}%
  \BibitemOpen
  \bibfield  {author} {\bibinfo {author} {\bibfnamefont {K.}~\bibnamefont
  {Nagata}}, \bibinfo {author} {\bibfnamefont {K.}~\bibnamefont {Kashiwa}},
  \bibinfo {author} {\bibfnamefont {A.}~\bibnamefont {Nakamura}}, \ and\
  \bibinfo {author} {\bibfnamefont {S.~M.}\ \bibnamefont {Nishigaki}},\ }\href
  {\doibase 10.1103/PhysRevD.91.094507} {\bibfield  {journal} {\bibinfo
  {journal} {Phys. Rev.}\ }\textbf {\bibinfo {volume} {D91}},\ \bibinfo {pages}
  {094507} (\bibinfo {year} {2015})},\ \Eprint {http://arxiv.org/abs/1410.0783}
  {arXiv:1410.0783 [hep-lat]} \BibitemShut {NoStop}%
\bibitem [{\citenamefont {de~Forcrand}\ and\ \citenamefont
  {Philipsen}(2002)}]{forcrand1}%
  \BibitemOpen
  \bibfield  {author} {\bibinfo {author} {\bibfnamefont {P.}~\bibnamefont
  {de~Forcrand}}\ and\ \bibinfo {author} {\bibfnamefont {O.}~\bibnamefont
  {Philipsen}},\ }\href {\doibase 10.1016/S0550-3213(02)00626-0} {\bibfield
  {journal} {\bibinfo  {journal} {Nucl.Phys.}\ }\textbf {\bibinfo {volume}
  {B642}},\ \bibinfo {pages} {290} (\bibinfo {year} {2002})},\ \Eprint
  {http://arxiv.org/abs/hep-lat/0205016} {arXiv:hep-lat/0205016 [hep-lat]}
  \BibitemShut {NoStop}%
\bibitem [{\citenamefont {de~Forcrand}\ and\ \citenamefont
  {Philipsen}(2003)}]{forcrand2}%
  \BibitemOpen
  \bibfield  {author} {\bibinfo {author} {\bibfnamefont {P.}~\bibnamefont
  {de~Forcrand}}\ and\ \bibinfo {author} {\bibfnamefont {O.}~\bibnamefont
  {Philipsen}},\ }\href {\doibase 10.1016/j.nuclphysb.2003.09.005} {\bibfield
  {journal} {\bibinfo  {journal} {Nucl.Phys.}\ }\textbf {\bibinfo {volume}
  {B673}},\ \bibinfo {pages} {170} (\bibinfo {year} {2003})},\ \Eprint
  {http://arxiv.org/abs/hep-lat/0307020} {arXiv:hep-lat/0307020 [hep-lat]}
  \BibitemShut {NoStop}%
\bibitem [{\citenamefont {D'Elia}\ and\ \citenamefont
  {Lombardo}(2003)}]{delia1}%
  \BibitemOpen
  \bibfield  {author} {\bibinfo {author} {\bibfnamefont {M.}~\bibnamefont
  {D'Elia}}\ and\ \bibinfo {author} {\bibfnamefont {M.-P.}\ \bibnamefont
  {Lombardo}},\ }\href {\doibase 10.1103/PhysRevD.67.014505} {\bibfield
  {journal} {\bibinfo  {journal} {Phys.Rev.}\ }\textbf {\bibinfo {volume}
  {D67}},\ \bibinfo {pages} {014505} (\bibinfo {year} {2003})},\ \Eprint
  {http://arxiv.org/abs/hep-lat/0209146} {arXiv:hep-lat/0209146 [hep-lat]}
  \BibitemShut {NoStop}%
\bibitem [{\citenamefont {D'Elia}\ and\ \citenamefont
  {Lombardo}(2004)}]{delia2}%
  \BibitemOpen
  \bibfield  {author} {\bibinfo {author} {\bibfnamefont {M.}~\bibnamefont
  {D'Elia}}\ and\ \bibinfo {author} {\bibfnamefont {M.~P.}\ \bibnamefont
  {Lombardo}},\ }\href {\doibase 10.1103/PhysRevD.70.074509} {\bibfield
  {journal} {\bibinfo  {journal} {Phys.Rev.}\ }\textbf {\bibinfo {volume}
  {D70}},\ \bibinfo {pages} {074509} (\bibinfo {year} {2004})},\ \Eprint
  {http://arxiv.org/abs/hep-lat/0406012} {arXiv:hep-lat/0406012 [hep-lat]}
  \BibitemShut {NoStop}%
\bibitem [{\citenamefont {Chen}\ and\ \citenamefont {Luo}(2005)}]{chen}%
  \BibitemOpen
  \bibfield  {author} {\bibinfo {author} {\bibfnamefont {H.-S.}\ \bibnamefont
  {Chen}}\ and\ \bibinfo {author} {\bibfnamefont {X.-Q.}\ \bibnamefont {Luo}},\
  }\href {\doibase 10.1103/PhysRevD.72.034504} {\bibfield  {journal} {\bibinfo
  {journal} {Phys.Rev.}\ }\textbf {\bibinfo {volume} {D72}},\ \bibinfo {pages}
  {034504} (\bibinfo {year} {2005})},\ \Eprint
  {http://arxiv.org/abs/hep-lat/0411023} {arXiv:hep-lat/0411023 [hep-lat]}
  \BibitemShut {NoStop}%
\bibitem [{\citenamefont {Schnetz}\ \emph {et~al.}(2006)\citenamefont
  {Schnetz}, \citenamefont {Thies},\ and\ \citenamefont {Urlichs}}]{schnetz}%
  \BibitemOpen
  \bibfield  {author} {\bibinfo {author} {\bibfnamefont {O.}~\bibnamefont
  {Schnetz}}, \bibinfo {author} {\bibfnamefont {M.}~\bibnamefont {Thies}}, \
  and\ \bibinfo {author} {\bibfnamefont {K.}~\bibnamefont {Urlichs}},\ }\href
  {\doibase http://dx.doi.org/10.1016/j.aop.2005.12.007} {\bibfield  {journal}
  {\bibinfo  {journal} {Annals of Physics}\ }\textbf {\bibinfo {volume}
  {321}},\ \bibinfo {pages} {2604 } (\bibinfo {year} {2006})}\BibitemShut
  {NoStop}%
\bibitem [{\citenamefont {Karasawa}\ and\ \citenamefont
  {Tatsumi}(2013)}]{karasawa}%
  \BibitemOpen
  \bibfield  {author} {\bibinfo {author} {\bibfnamefont {S.}~\bibnamefont
  {Karasawa}}\ and\ \bibinfo {author} {\bibfnamefont {T.}~\bibnamefont
  {Tatsumi}},\ }\href@noop {} {\bibfield  {journal} {\bibinfo  {journal}
  {arXiv:1309.6448}\ } (\bibinfo {year} {2013})}\BibitemShut {NoStop}%
\bibitem [{\citenamefont {Kashiwa}\ \emph {et~al.}(2015)\citenamefont
  {Kashiwa}, \citenamefont {Lee}, \citenamefont {Nishiyama},\ and\
  \citenamefont {Yoshiike}}]{kashiwa}%
  \BibitemOpen
  \bibfield  {author} {\bibinfo {author} {\bibfnamefont {K.}~\bibnamefont
  {Kashiwa}}, \bibinfo {author} {\bibfnamefont {T.-G.}\ \bibnamefont {Lee}},
  \bibinfo {author} {\bibfnamefont {K.}~\bibnamefont {Nishiyama}}, \ and\
  \bibinfo {author} {\bibfnamefont {R.}~\bibnamefont {Yoshiike}},\ }\href@noop
  {} {\  (\bibinfo {year} {2015})},\ \Eprint {http://arxiv.org/abs/1507.08382}
  {arXiv:1507.08382 [hep-ph]} \BibitemShut {NoStop}%
\bibitem [{\citenamefont {Klevansky}(1992)}]{njl}%
  \BibitemOpen
  \bibfield  {author} {\bibinfo {author} {\bibfnamefont {S.~P.}\ \bibnamefont
  {Klevansky}},\ }\href {\doibase 10.1103/RevModPhys.64.649} {\bibfield
  {journal} {\bibinfo  {journal} {Rev. Mod. Phys.}\ }\textbf {\bibinfo {volume}
  {64}},\ \bibinfo {pages} {649} (\bibinfo {year} {1992})}\BibitemShut
  {NoStop}%
\bibitem [{\citenamefont {Ferreira}\ \emph {et~al.}(2014)\citenamefont
  {Ferreira}, \citenamefont {Costa}, \citenamefont
  {Louren\ifmmode~\mbox{\c{c}}\else \c{c}\fi{}o}, \citenamefont {Frederico},\
  and\ \citenamefont {Provid\^encia}}]{ferreira}%
  \BibitemOpen
  \bibfield  {author} {\bibinfo {author} {\bibfnamefont {M.}~\bibnamefont
  {Ferreira}}, \bibinfo {author} {\bibfnamefont {P.}~\bibnamefont {Costa}},
  \bibinfo {author} {\bibfnamefont {O.}~\bibnamefont
  {Louren\ifmmode~\mbox{\c{c}}\else \c{c}\fi{}o}}, \bibinfo {author}
  {\bibfnamefont {T.}~\bibnamefont {Frederico}}, \ and\ \bibinfo {author}
  {\bibfnamefont {C.}~\bibnamefont {Provid\^encia}},\ }\href {\doibase
  10.1103/PhysRevD.89.116011} {\bibfield  {journal} {\bibinfo  {journal} {Phys.
  Rev. D}\ }\textbf {\bibinfo {volume} {89}},\ \bibinfo {pages} {116011}
  (\bibinfo {year} {2014})}\BibitemShut {NoStop}%
\bibitem [{\citenamefont {Kogut}\ \emph {et~al.}(2001)\citenamefont {Kogut},
  \citenamefont {Sinclair}, \citenamefont {Hands},\ and\ \citenamefont
  {Morrison}}]{kogut}%
  \BibitemOpen
  \bibfield  {author} {\bibinfo {author} {\bibfnamefont {J.~B.}\ \bibnamefont
  {Kogut}}, \bibinfo {author} {\bibfnamefont {D.~K.}\ \bibnamefont {Sinclair}},
  \bibinfo {author} {\bibfnamefont {S.~J.}\ \bibnamefont {Hands}}, \ and\
  \bibinfo {author} {\bibfnamefont {S.~E.}\ \bibnamefont {Morrison}},\ }\href
  {\doibase 10.1103/PhysRevD.64.094505} {\bibfield  {journal} {\bibinfo
  {journal} {Phys. Rev. D}\ }\textbf {\bibinfo {volume} {64}},\ \bibinfo
  {pages} {094505} (\bibinfo {year} {2001})}\BibitemShut {NoStop}%
\bibitem [{\citenamefont {Kashiwa}\ \emph {et~al.}()\citenamefont {Kashiwa},
  \citenamefont {Lee},\ and\ \citenamefont {Yoshiike}}]{kashiwa2}%
  \BibitemOpen
  \bibfield  {author} {\bibinfo {author} {\bibfnamefont {K.}~\bibnamefont
  {Kashiwa}}, \bibinfo {author} {\bibfnamefont {T.-G.}\ \bibnamefont {Lee}}, \
  and\ \bibinfo {author} {\bibfnamefont {R.}~\bibnamefont {Yoshiike}},\
  }\href@noop {} {\bibinfo  {journal} {in preparation}\ }\BibitemShut {NoStop}%
\bibitem [{\citenamefont {Iritani}\ \emph {et~al.}(2015)\citenamefont
  {Iritani}, \citenamefont {Cossu},\ and\ \citenamefont {Hashimoto}}]{iritani}%
  \BibitemOpen
\bibfield  {journal} {  }\bibfield  {author} {\bibinfo {author} {\bibfnamefont
  {T.}~\bibnamefont {Iritani}}, \bibinfo {author} {\bibfnamefont
  {G.}~\bibnamefont {Cossu}}, \ and\ \bibinfo {author} {\bibfnamefont
  {S.}~\bibnamefont {Hashimoto}},\ }\href {\doibase 10.1103/PhysRevD.91.094501}
  {\bibfield  {journal} {\bibinfo  {journal} {Phys. Rev. D}\ }\textbf {\bibinfo
  {volume} {91}},\ \bibinfo {pages} {094501} (\bibinfo {year}
  {2015})}\BibitemShut {NoStop}%
\bibitem [{\citenamefont {Moreira}\ \emph {et~al.}(2014)\citenamefont
  {Moreira}, \citenamefont {Hiller}, \citenamefont {Broniowski}, \citenamefont
  {Osipov},\ and\ \citenamefont {Blin}}]{moreira}%
  \BibitemOpen
  \bibfield  {author} {\bibinfo {author} {\bibfnamefont {J.}~\bibnamefont
  {Moreira}}, \bibinfo {author} {\bibfnamefont {B.}~\bibnamefont {Hiller}},
  \bibinfo {author} {\bibfnamefont {W.}~\bibnamefont {Broniowski}}, \bibinfo
  {author} {\bibfnamefont {A.~A.}\ \bibnamefont {Osipov}}, \ and\ \bibinfo
  {author} {\bibfnamefont {A.~H.}\ \bibnamefont {Blin}},\ }\href {\doibase
  10.1103/PhysRevD.89.036009} {\bibfield  {journal} {\bibinfo  {journal} {Phys.
  Rev. D}\ }\textbf {\bibinfo {volume} {89}},\ \bibinfo {pages} {036009}
  (\bibinfo {year} {2014})}\BibitemShut {NoStop}%
\bibitem [{\citenamefont {Carignano}\ \emph {et~al.}(2015)\citenamefont
  {Carignano}, \citenamefont {Ferrer},\ and\ \citenamefont {Incela}}]{eos}%
  \BibitemOpen
  \bibfield  {author} {\bibinfo {author} {\bibfnamefont {S.}~\bibnamefont
  {Carignano}}, \bibinfo {author} {\bibfnamefont {E.~J.}\ \bibnamefont
  {Ferrer}}, \ and\ \bibinfo {author} {\bibfnamefont {V.~d.~l.}\ \bibnamefont
  {Incela}},\ }\href@noop {} {\  (\bibinfo {year} {2015})},\ \Eprint
  {http://arxiv.org/abs/1505.05094} {arXiv:1505.05094 [nucl-th]} \BibitemShut
  {NoStop}%
\bibitem [{\citenamefont {Hidaka}\ \emph {et~al.}(2015)\citenamefont {Hidaka},
  \citenamefont {Kamikado}, \citenamefont {Kanazawa},\ and\ \citenamefont
  {Noumi}}]{kamikado}%
  \BibitemOpen
  \bibfield  {author} {\bibinfo {author} {\bibfnamefont {Y.}~\bibnamefont
  {Hidaka}}, \bibinfo {author} {\bibfnamefont {K.}~\bibnamefont {Kamikado}},
  \bibinfo {author} {\bibfnamefont {T.}~\bibnamefont {Kanazawa}}, \ and\
  \bibinfo {author} {\bibfnamefont {T.}~\bibnamefont {Noumi}},\ }\href
  {\doibase 10.1103/PhysRevD.92.034003} {\bibfield  {journal} {\bibinfo
  {journal} {Phys. Rev.}\ }\textbf {\bibinfo {volume} {D92}},\ \bibinfo {pages}
  {034003} (\bibinfo {year} {2015})},\ \Eprint
  {http://arxiv.org/abs/1505.00848} {arXiv:1505.00848 [hep-ph]} \BibitemShut
  {NoStop}%
\bibitem [{\citenamefont {Schwinger}(1951)}]{schwinger}%
  \BibitemOpen
  \bibfield  {author} {\bibinfo {author} {\bibfnamefont {J.}~\bibnamefont
  {Schwinger}},\ }\href {\doibase 10.1103/PhysRev.82.664} {\bibfield  {journal}
  {\bibinfo  {journal} {Phys. Rev.}\ }\textbf {\bibinfo {volume} {82}},\
  \bibinfo {pages} {664} (\bibinfo {year} {1951})}\BibitemShut {NoStop}%
\bibitem [{\citenamefont {Chodos}\ \emph {et~al.}(1990)\citenamefont {Chodos},
  \citenamefont {Everding},\ and\ \citenamefont {Owen}}]{chodos}%
  \BibitemOpen
  \bibfield  {author} {\bibinfo {author} {\bibfnamefont {A.}~\bibnamefont
  {Chodos}}, \bibinfo {author} {\bibfnamefont {K.}~\bibnamefont {Everding}}, \
  and\ \bibinfo {author} {\bibfnamefont {D.~A.}\ \bibnamefont {Owen}},\ }\href
  {\doibase 10.1103/PhysRevD.42.2881} {\bibfield  {journal} {\bibinfo
  {journal} {Phys. Rev. D}\ }\textbf {\bibinfo {volume} {42}},\ \bibinfo
  {pages} {2881} (\bibinfo {year} {1990})}\BibitemShut {NoStop}%
\bibitem [{\citenamefont {Niemi}\ and\ \citenamefont {Semenoff}(1986)}]{niemi}%
  \BibitemOpen
  \bibfield  {author} {\bibinfo {author} {\bibfnamefont {A.~J.}\ \bibnamefont
  {Niemi}}\ and\ \bibinfo {author} {\bibfnamefont {G.~W.}\ \bibnamefont
  {Semenoff}},\ }\href@noop {} {\bibfield  {journal} {\bibinfo  {journal}
  {Phys. Rep.}\ }\textbf {\bibinfo {volume} {135}},\ \bibinfo {pages} {99}
  (\bibinfo {year} {1986})}\BibitemShut {NoStop}%
\bibitem [{\citenamefont {Niemi}(1985)}]{niemi2}%
  \BibitemOpen
  \bibfield  {author} {\bibinfo {author} {\bibfnamefont {A.~J.}\ \bibnamefont
  {Niemi}},\ }\href@noop {} {\bibfield  {journal} {\bibinfo  {journal} {Nucl.
  Phys. B}\ }\textbf {\bibinfo {volume} {251}},\ \bibinfo {pages} {155}
  (\bibinfo {year} {1985})}\BibitemShut {NoStop}%
\bibitem [{\citenamefont {Son}\ and\ \citenamefont {Stephanov}(2008)}]{son}%
  \BibitemOpen
  \bibfield  {author} {\bibinfo {author} {\bibfnamefont {D.~T.}\ \bibnamefont
  {Son}}\ and\ \bibinfo {author} {\bibfnamefont {M.~A.}\ \bibnamefont
  {Stephanov}},\ }\href@noop {} {\bibfield  {journal} {\bibinfo  {journal}
  {Phys. Rev. D}\ }\textbf {\bibinfo {volume} {77}},\ \bibinfo {pages} {014021}
  (\bibinfo {year} {2008})}\BibitemShut {NoStop}%
\end{thebibliography}%

\end{document}